\documentclass[12pt]{article}
\pdfoutput=1

\usepackage{amsmath,amssymb,amscd}
\usepackage{listings}
\usepackage{caption}
\usepackage{dsfont}
\usepackage{slashed}
\usepackage{color}

\usepackage[pdftex]{graphicx}
\usepackage{epstopdf}
\usepackage{subfigure}
\usepackage{epsfig}
\usepackage{listings}
\usepackage{caption}
\usepackage{cite}

\usepackage{multirow}
\usepackage[titletoc]{appendix}

\usepackage{array}
\newcolumntype{M}[1]{>{\centering\arraybackslash}m{#1}}
\newcolumntype{N}{@{}m{0pt}@{}}

\newcommand{\bea}{\begin{align}}
\newcommand{\eea}{\end{align}}
\newcommand{\beq}{\begin{equation}}
\newcommand{\eeq}{\end{equation}}
\newcommand{\nbea}{\begin{align*}}
\newcommand{\neea}{\end{align*}}
\newcommand{\nbeq}{\begin{equation*}}
\newcommand{\neeq}{\end{equation*}}
\newcommand{\bear}{\begin{eqnarray}}  
\newcommand{\eear}{\end{eqnarray}}  

 \newcommand{\twomatrix}[1]{\left(\begin{array}{cc} #1 \end{array}\right) }

 \newcommand{\identity}{\mathds{1}}

\newcommand{\cc}[2]{\ensuremath{{}_{#1}^{#2}}}
\newcommand{\I}{\text{I}}
\newcommand{\Id}{\text{I}[\text{q}^{2}]}
\newcommand{\Iq}{\text{I}[\text{q}^{4}]}

\def\abs#1{\left| #1 \right|}
\newcommand{\Dfbu}{\mathord{\buildrel{\lower3pt\hbox{$\scriptscriptstyle{\leftrightarrow \tiny{ \ \ \ } }$}}\over {D^{\mu}}}} 
\newcommand{\Dfbd}{\mathord{\buildrel{\lower3pt\hbox{$\scriptscriptstyle\leftrightarrow$}}\over {D}_{\mu}}} 

\numberwithin{equation}{section}

\begin{document}

\begin{titlepage}

\pagestyle{empty}

\baselineskip=21pt
\rightline{\small KCL-PH-TH/2015-54, LCTS/2015-42, CERN-PH-TH/2015-284}
\rightline{\small Cavendish-HEP-15/12, DAMTP-2015-88}
\vskip 0.8in

\begin{center}

{\large {\bf The Universal One-Loop Effective Action}}

\vskip 0.6in

{\bf Aleksandra~Drozd}$^{1}$,
 {\bf John~Ellis}$^{1,2}$,
 {\bf J\'er\'emie~Quevillon}$^{1}$
and {\bf Tevong~You}$^{1,3}$

\vskip 0.4in

{\small {\it
$^1${Theoretical Particle Physics and Cosmology Group, Physics Department, \\
King's College London, London WC2R 2LS, UK}\\
\vspace{0.25cm}
$^2${TH Division, Physics Department, CERN, CH-1211 Geneva 23, Switzerland} \\
\vspace{0.25cm}
$^3${Cavendish Laboratory, University of Cambridge, J.J. Thomson Avenue, \\
Cambridge, CB3 0HE, UK; \\
\vspace{-0.25cm}
DAMTP, University of Cambridge, Wilberforce Road, Cambridge, CB3 0WA, UK}
}}

\vskip 0.75in

{\bf Abstract}

\end{center}

\baselineskip=18pt \noindent


{\small

We present the universal one-loop effective action for all operators of dimension up to six obtained by integrating out
massive, non-degenerate multiplets. Our general expression may be applied to loops of heavy fermions or bosons,
and has been checked against partial results available in the literature. The broad applicability of this approach simplifies 
one-loop matching from an ultraviolet model to a lower-energy effective field theory (EFT), a procedure which is now reduced to the 
evaluation of a combination of matrices in our universal expression, without any loop integrals to evaluate. We illustrate the 
relationship of our results to the Standard Model (SM) EFT, using as an example the supersymmetric stop and sbottom squark Lagrangian
and extracting from our universal expression the Wilson coefficients of dimension-six operators composed of SM fields.  
}


\vskip 1in

{\small \leftline{December 2015}}

\end{titlepage}

\newpage


\section{Introduction}

Following the discovery of the Higgs boson at the LHC~\cite{Higgs}, and in the absence so far
of any evidence at the LHC for physics beyond the Standard Model (SM), attention is focusing on possible
indirect signatures of new massive particles. These could manifest themselves either via tree-level
exchanges, much as the the first indirect evidence for the $W$ boson came from the four-fermion
weak interaction, or via loop effects, which was how the first indirect evidence was found for the $c$
and $t$ quarks and the Higgs boson. In those cases, loop calculations also provided
indications on their possible masses.

Matching from an ultraviolet (UV) theory to a low-energy effective field theory (EFT) can be achieved by calculating 
Feynman diagrams in the UV and EFT, or by evaluating the path integral to integrate out directly the heavy particles. 
A manifestly gauge-invariant method for simplifying calculations in the latter approach is the covariant derivative expansion (CDE) 
pioneered by Gaillard in~\cite{Gaillard} and by Cheyette in~\cite{Cheyette}. The Gaillard-Cheyette CDE was recently revived by 
Henning, Lu and Murayama (HLM)~\cite{HLM}. It was noticed by HLM that under the assumption of degenerate masses for the 
heavy multiplets connected by off-diagonal entries in the interaction matrix of the quadratic term, the momentum integrals factored out of the calculation and could be evaluated independently of the details of the UV theory 
and the specific field content of the resulting EFT. This universality property enabled them to obtain a one-loop effective action that applies quite generally, 
albeit under the restrictive condition that the particles in the UV theory are degenerate in mass~\cite{HLM}. 

The main purpose of this paper is to present general formulae for the terms in the effective one-loop action due to the exchanges of massive virtual particles, 
bosons or fermions, that do {\it not} appear in degenerate multiplets when these interact with each other. As we show, it is possible to generalize the HLM results 
and write the relevant one-loop effective action in a universal closed form without any assumptions on the mass spectrum and the form of the quadratic interaction matrix. 
We write down the complete expression for the universal one-loop effective action relevant for obtaining the Wilson coefficients of 
all operators up to dimension six in the EFT. We also provide details on how this result was obtained using the CDE, 
which can in principle be used to extend the universal one-loop effective action to higher-dimensional operators. 

These general results have immediate phenomenological applicability to the interpretation of
precision SM measurements. With all the SM particle content now experimentally established, 
a popular way of encapsulating the possible indirect effects of new massive particles is via
an EFT composed of SM fields~\cite{buchmullerwyler, GIMR}. This SM EFT approach is a
powerful and systematic tool for capturing the effects of virtual massive particles in the decoupling limit~\cite{decoupling},
and very convenient for implementing and combining the experimental constraints from 
different classes of measurements in a consistent way~\footnote{See for example Refs.~\cite{earlyeft} and \cite{recenteftpapers, mixedlightheavy, bilenkysantamaria, habaetal, craigetal, gorbahnetal, HLMstop, DEQY1, NLO, RHvectorlike, RanHuo, EYfuture, wellszhanguniversalEFT} for a sampling of early and recent studies, and \cite{SMEFTreviews} for reviews.}. In particular, the SM EFT operators
of dimension 6 that are invariant under the SM SU(3) $\times$ SU(2) $\times$ U(1) gauge group
have been classified~\cite{buchmullerwyler}, suitable non-redundant bases of independent operators
have been identified~\cite{eomreduction, GIMR, SILH, rosetta}, the anomalous dimension matrix for RG running has been derived~\cite{dim6RGE}, and global analyses of the possible magnitudes of their coefficients
have been undertaken for present~\cite{HanSkiba, eftconstraints, ciuchinietal, PomarolRiva, ESY, falkowskiriva, berthiertrott, flavourfulEFT, higgslegacyEFT, topquarkEFT, TGCEFT, brehmeretal, englertetal,falkowskimimouni} and future colliders~\cite{craigetal, EYfuture}. The universal one-loop effective action that
we present facilitates making contact with specific UV models. 

We illustrate this procedure in the minimally supersymmetric SM (MSSM). Starting from the Lagrangian of the third-generation 
left-handed squark doublet and the right-handed stop, we pick out the relevant matrices of SM fields and gauge field strengths
to plug into the universal one-loop expression. The Wilson coefficients for heavy stops are then automatically obtained, 
with the rearrangement of the SM fields in the desired operator basis as the only non-trivial step. No one-loop calculations of 
momentum integrals have to be performed, as these are already encapsulated in the coefficients of the universal expression.

Our general result allows the supersymmetry-breaking masses for the left-handed squark doublet and the right-handed stop 
to be non-degenerate. We also allow mass splittings within multiplets, such as those due to electroweak symmetry breaking,
which are generated by the non-zero vacuum expectation value $v$ of the Englert-Brout-Higgs
field. If the massive particles are sufficiently heavy, the effects of non-degeneracy are
suppressed by powers of $v$ divided by the heavy mass scale. However, the present
constraints on the loop effects due to massive particles are not sufficiently strong for
their possible non-degeneracy to be insignificant. A case in point is provided by
stop and sbottom squarks, some of which could well have masses that are not much
larger than $v$. These effects have been studied in Ref.~\cite{craigetal, DEQY1} and will not be considered further here, 
but they could in principle be captured by our universal one-loop effective action, as it accommodates such splittings within multiplets. 

We would like to emphasize the advantages of the Gaillard-Cheyette CDE method that we used to derive the universal one-loop effective action in this paper.
It is {\it manifestly gauge-invariant}, which determines directly the form of the universal effective action, it {\it is not restricted to any particular approximation}, 
though the series expansion may be evaluated up to some fixed order as we have done to obtain our universal one-loop action, 
and it may be used to explore effects due to {\it any} UV theory or extension of the SM. These advantages translate to the universal one-loop effective action,
which may be evaluated to any fixed order, though here we restrict our calculation to all terms in the CDE series expansion relevant for obtaining operators of dimension up to six. 

We have performed various cross-checks on our calculations, including checks against
partial results available in the literature. In particular, we reproduce using our universal expression
the HLM results for the universal one-loop effective action in the degenerate case~\cite{HLM}, 
and the contributions of degenerate squarks to the coefficients of bosonic Higgs operators obtained in~\cite{HLMstop}. 
In the non-degenerate case the Wilson coefficients for the squark case agrees in part with those in~\cite{RanHuo}, 
and we comment on the differences with our expressions, which are partially but not entirely due to differences in 
operator bases. As another independent check for the non-degenerate expressions, 
we find that the combination of squark Wilson coefficients corresponding to the Peskin-Takeuchi $S$ and $T$ parameters~\cite{PeskinTakeuchi} (or equivalently the Altarelli-Barbieri $\epsilon_{1,2}$ parameters~\cite{AltarelliBarbieri}) 
agree with past calculations of $S$ and $T$ in the MSSM. 

The structure of this paper is as follows. In Section~\ref{sec:universaleffectiveaction} we present the universal one-loop effective action 
and its relation to the general form of the UV model being integrated out, showing that it reproduces the HLM result in the degenerate limit. 
Our expression includes all terms needed for a complete one-loop matching of operators up to dimension six, 
though in principle such a universal one-loop effective action may be extended to include higher-dimensional operators by evaluating further
terms in the series expansion of the CDE. In Section~\ref{sec:CDE} we discuss the Gaillard-Cheyette CDE method for performing the series expansion 
of the logarithm in the path integral. This section is for the interested reader who may wish to double-check our results or extend them to higher-dimensional operators, 
or even bypass our universal one-loop expression entirely to integrate out heavy particles from beginning to end. 
This latter use is indeed how the path integral method has been employed until now, but we stress that for the purpose of calculating 
Wilson coefficients of operators up to dimension six  \emph{this is no longer necessary}, and the starting point should be to evaluate the 
sum over the indices in the universal one-loop effective action of Section~\ref{sec:universaleffectiveaction}. 
In Section~\ref{sec:conclusion} we summarise our general results and discuss some implications. Several calculational details and lengthy 
expressions for coefficients are described in Appendices~\ref{app:fcomputation}, \ref{app:masterintegrals} and \ref{app:fexplicit}. In Appendix~\ref{app:stop} we
apply our results to the concrete example of non-degenerate supersymmetric stop and sbottom squarks, checking against
previous calculations in the literature.

A {\tt Mathematica} notebook with explicit expressions for the universal coefficients and their integrals is available on request.

\section{The Universal One-Loop Effective Action}
\label{sec:universaleffectiveaction}

We consider a UV theory, renormalizable or not~\footnote{The general expressions we give later
can in fact be simplified in the renormalizable case for the SM EFT~\cite{inprep}.}, composed of light background fields, collectively denoted as $\phi$, and heavy fields arranged in a multiplet $\Phi$, which can be either fermions or bosons, with a generic Lagrangian consisting of a low-energy part plus the terms coupling the heavy fields to the light ones. This can be written as
\begin{equation}
\mathcal{L}_\text{UV}[\phi,\Phi] = \mathcal{L}[\phi] + (\Phi^\dagger F[\phi] + \text{h.c.}) + \Phi^\dagger(P^2 - M^2 - U[\phi])\Phi + \mathcal{O}(\Phi^3) \, ,
\label{eq:lagrangianUV}
\end{equation}
where we left out the kinetic and mass terms for $\Phi$, and introduced the notation $P \equiv iD_\mu$, with $D_\mu$ the gauge-covariant derivative. $F[\phi]$ and $U[\phi]$ are matrices involving combinations of light fields coupling linearly and quadratically respectively to $\Phi$, and $M$ is a diagonal mass matrix. The form of $U$ will depend on the scalar, vectorial or fermionic nature of $\Phi$ in order to write the Lagrangian in this way |  an example for the case of fermions is given in Appendix~\ref{app:fermions} and we refer to Ref.~\cite{HLM} for more details~\footnote{We assume here that $U$ does not contain any covariant derivatives. We thank the referee for pointing out that this may be of importance when applied to non-renormalisable theories, which we leave for future studies.}.

For example, $F = \eta|H|^2$ if $\Phi$ is a heavy real singlet scalar coupling to the light Higgs doublet $H$ through $\mathcal{L} \supset \eta \Phi |H|^2$, or $U = \eta_1 |H|^2 + \eta_2 {\tilde H}{\tilde H}^\dagger$ if $\Phi$ is a scalar electroweak doublet with hypercharge $Y = -1/2$ and with a global $U(1)$ symmetry to restrict the Lagrangian to quadratic couplings of the form
 \begin{equation}
 \mathcal{L} \supset \Phi^\dagger( -\eta_1 |H|^2 - \eta_2 {\tilde H}{\tilde H}^\dagger)\Phi \, . 
 \label{eq:ewdoubletexample}
 \end{equation}
 In more complicated models there may be interactions between several heavy multiplets one wishes to integrate out, so $U$ is in general a matrix. 

The path integral over $\Phi$ may be computed by expanding the action 
around the minimum with respect to $\Phi$. The terms linear in the heavy fields involving $F[\phi]$ then enter in the tree-level effective Lagrangian upon substituting $\Phi_c$, the solution to the equation of motion, which gives~\cite{HLM}
\begin{equation*}
\mathcal{L}^\text{eff}_\text{tree}[\phi] = \sum_{n=0}^\infty F^\dagger M^{-2} [(P^2-U)M^{-2}]^n F + \mathcal{O}(\Phi_c^3) \, .
\end{equation*}
The terms quadratic in the heavy fields are responsible for the one-loop part of the effective Lagrangian and can be evaluated in the path integral using the CDE method described in Section~\ref{sec:CDE}. We simply state here the final expression, which is the main result of this paper:
\begin{align}
\nonumber \\
{\cal L}^{\text{eff}}_{\text{1-loop}}[\phi] \supset -i c_s & \Bigg\{ 
f_{1}^{i} + f_{2}^{i}U_{ii} + f_{3}^{i} G_{\mu\nu, ij}'^2 + f_{4}^{ij} U_{ij}^2  \nonumber \\
& + f_{5}^{ij}(P_{\mu} G_{\mu \nu,ij}^{'})^{2} + f_{6}^{ij}(G_{\mu \nu,ij}^{'})(G_{\nu \sigma,jk}^{'}) (G_{\sigma \mu,ki}^{'}) +f_{7}^{ij} [P_{\mu}, U_{ij}]^2 +f_{8}^{ijk} (U_{ij}U_{jk}U_{ki})   
\nonumber \\
& 
+ f_{9}^{ij} (U_{ij} G_{\mu\nu,jk}'G_{\mu\nu,ki}^{'}) \nonumber \\
&+ f_{10}^{ijkl} (U_{ij}U_{jk}U_{kl} U_{li}) + f_{11}^{ijk} U_{ij} [P_{\mu},U_{jk}][P_{\mu},U_{ki}]  \nonumber \\
& +f_{12,a}^{ij} \left[P_{\mu},[P_{\nu},U_{ij}]\right]\left[P_{\mu}, [P_{\nu},U_{ji}]\right] + f_{12,b}^{ij} \left[P_{\mu},[P_{\nu},U_{ij}]\right] \left[ P_{\nu}, [P_{\mu},U_{ji}]\right]  \nonumber \\
& + f_{12,c}^{ij} \left[P_{\mu},[P_{\mu},U_{ij}]\right]\left[P_{\nu},[P_{\nu},U_{ji}]\right]
\nonumber \\
& +f_{13}^{ijk} U_{ij} U_{jk} G_{\mu \nu,kl}^{'}G_{\mu \nu,li}^{'} + f_{14}^{ijk}\left[ P_{\mu},U_{ij}\right] \left[ P_{\nu},U_{jk}\right] G_{\nu \mu,ki}^{'}  \nonumber \\
& +\left(f_{15a}^{ijk} U_{i,j} [ P_{\mu},U_{j,k}] - f_{15b}^{ijk} [P_{\mu},U_{i,j}] U_{j,k} \right) [P_{\nu},G^{'}_{\nu\mu , ki}] \nonumber \\
& + f_{16}^{ijklm}(U_{ij}U_{jk}U_{kl} U_{lm} U_{mi}) + f_{17}^{ijkl} U_{ij} U_{jk}  [P_{\mu},U_{kl}][P_{\mu},U_{li}] +f_{18}^{ijkl}  U_{ij} [P_{\mu},U_{jk}] U_{kl} [P_{\mu},U_{li}]
\nonumber \\
&   +f_{19}^{ijklmn} (U_{ij}U_{jk}U_{kl} U_{lm} U_{mn} U_{ni}) 
 \Bigg\} \, .
 \nonumber \\
\label{eq:universallagrangian}
\end{align}
The constant $c_{s} = 1/2,1,-1/2,1/2$ and $-1$ when integrating out real scalars, complex scalars, 
Dirac fermions, gauge bosons and Fadeev-Popov ghosts respectively. The indices $i,j,k,l$ range over the dimension of the mass matrix $M$, with an implied summation when the same index appears twice in a single term. We have introduced a notation for the field strength matrix $G^\prime_{\mu\nu} \equiv -igG_{\mu\nu}$, where $g$ is the gauge coupling for each specific field strength within the matrix $G_{\mu\nu}$~\footnote{This notation is figurative and not to be taken literally, as the coupling does not factor out of the matrix, in general. The field strength matrix is related to the covariant derivative by $[P_\mu,P_\nu] = -G^\prime_{\mu\nu}$, which may in general contain several gauge fields and couplings, e.g., $G^\prime_{\mu\nu} = -igW^a_{\mu\nu}\tau^a-ig^\prime YB_{\mu\nu}\identity$ for the electroweak scalar doublet example of Eq.~(\ref{eq:ewdoubletexample}). }. Finally the $f_{N}^{ijk\cdots}$ are universal coefficients containing the mass parameter dependence that we describe shortly.

 A few comments about Eq.~(\ref{eq:universallagrangian}) are in order. We note that no specific UV theory has been specified or assumptions made, other than the general form of Eq.~\ref{eq:lagrangianUV}, so that this result holds model-independently. It may be used to obtain an EFT involving both external bosons and fermions~\footnote{To obtain a one-loop effective action with light external fermions requires integrating out the heavy scalars and fermions to which they couple, which may not be performed simultaneously as a single quadratic term in the path integral. However, this can be done by first integrating out at tree level the heavy fermion to obtain an effective quadratic term for the scalar, or vice versa. See for example Ref.~\cite{habaetal}.  }. The details of the UV model are encapsulated in the matrix $U$, the matrix $G^\prime_{\mu\nu}$, and the covariant derivative $P_\mu$, which are all functions of the low-energy degrees of freedom $\phi$ (which we recall collectively denotes any bosons or fermions in the EFT). Since $U$ has at least mass-dimension 1 we see that all the terms required for computing operators of dimension up to six are present in Eq.~\ref{eq:universallagrangian}. We have organised the Lagrangian so that the first line contains all the operators of dimension up to four. In the remaining lines there can also be contributions to operators with dimension $> 6$ in Eq.~\ref{eq:universallagrangian}, but these will not be completely accounted for by our expression since higher order terms in the CDE series expansion that we have truncated will also contribute to these operators. There is however nothing limiting the extension of our universal results to operators of dimension seven~\cite{dim7} and higher~\cite{dim8}.  

The universal coefficients $f_{N}^{ijk\cdots}$  are obtained by Feynman integrals over momentum, and are universal in the sense that they factor out of the UV-dependent matrix evaluations and have only to be computed once and for all. This universality was previously noticed by HLM in the restricted case of $M$ commuting with $U$ and $G^\prime_{\mu\nu}$~\cite{HLM}, and we have shown here that this property holds in the fully general case without any such degenerate mass assumptions. This clears the way for a simpler alternative to computing Feynman diagrams or evaluating path integrals for one-loop matching: One may now start directly from Eq.~\ref{eq:universallagrangian} with only the summation over the matrix indices left to perform for each specific model, thus avoiding redundant effort in the previous steps.   

The master integrals that enter in the universal coefficients are defined as
\beq
\text{I}[\text{q}^{2\alpha}]\cc{i}{n}\cc{j}{m}\cc{\cdots}{\cdots}\cc{l}{p} = \int \frac{d^4 \text{q}}{(2 \pi)^4} \int d\xi \,   \text{q}^{2\alpha}\left(\Delta_{\xi,i} \right)^n \left(\Delta_{\xi,j} \right)^m \cdots \left(\Delta_{\xi,l} \right)^p \, ,
\label{def1}
\eeq
where $\Delta_{\xi,i} = 1/(q^2 - \xi m^2_i)$ and $m_i \equiv M_{ii}$. We provide details for the derivation of the universal coefficients $f_{N}^{ijk\cdots}$ in Appendix~\ref{app:fcomputation}, the results of which can be written as follows:
{\small
\begin{eqnarray}
 f_{1}^{i} &=& \I \cc{i}{1} m_{i}^{2} \, , \nonumber \\
 f_{2}^{i} &=& \I \cc{i}{2} m_{i}^{2} \, , \nonumber \\
 f_{3}^{i} &=&\frac{1}{2}\left(\I \cc{i}{3} - \frac{4}{d}\Id \cc{i}{4} \right) m_{i}^{2} \, , \nonumber \\
 f_{4}^{ij} &=& \I \cc{i}{2} \cc{j}{1} m_{i}^{2} \, , \nonumber \\
 f_{5}^{ij} &=&  \frac{1}{9} \left(-\Id \cc{i}{3}\cc{j}{2}+2 \I\cc{i}{3}\cc{j}{1}-28 \Iq\cc{i}{6}\cc{j}{0}-2\Id\cc{i}{4}\cc{j}{1}+42 \Id\cc{i}{5}\cc{j}{0}-14 \I\cc{i}{4}\cc{j}{0}\right) m_i^2+\left(4 \Iq\cc{i}{0}\cc{j}{6}-6 \Id\cc{i}{0}\cc{j}{5}+ 2\I\cc{i}{0}\cc{j}{4}\right) m_j^2 \, , \nonumber \\
 f_{6}^{ij} &=& -\frac{7}{3} \left(2 \Iq\cc{i}{6}\cc{j}{0}-3 \Id\cc{i}{5}\cc{j}{0}+\I\cc{i}{4}\cc{j}{0}\right) m_i^2 +3 \left(2 \Iq\cc{i}{0}\cc{j}{6}-3 \Id\cc{i}{0}\cc{j}{5}+\I\cc{i}{0}\cc{j}{4}\right) m_j^2 \, , \nonumber \\
 f_{7}^{ij} &=& \left(\I\cc{i}{2}\cc{j}{2}-\Id\cc{i}{2}\cc{j}{3}\right) m_{i}^{2} \, , \nonumber \\
 \nonumber
    \end{eqnarray}
 \begin{eqnarray}
 f_{8}^{ijk} &=& \I\cc{i}{2}\cc{j}{1}\cc{k}{1} m_{i}^{2} \, , \nonumber \\
 f_{9}^{ij} &=& \frac{1}{2} \Big(\left(\I\cc{i}{3}\cc{j}{1}-\Id\cc{i}{4}\cc{j}{1}\right) m_i^2+\left(-\Id\cc{i}{1}\cc{j}{4}-\Id\cc{i}{2}\cc{j}{3}-\Id\cc{i}{3}\cc{j}{2}+\I\cc{i}{1}\cc{j}{3}+\I\cc{i}{2}\cc{j}{2}\right) m_j^2 \Big) \, , \nonumber \\
 f_{10}^{ijkl} &=& \I\cc{i}{2}\cc{j}{1}\cc{k}{1}\cc{l}{1} m_{i}^{2} \, , \nonumber \\ 
 f_{11}^{ijk} &=& \left(\I\cc{i}{2}\cc{j}{1}\cc{k}{2}-\Id\cc{i}{2}\cc{j}{1}\cc{k}{3}\right) m_i^2+\left(\I\cc{i}{1}\cc{j}{2}\cc{k}{2}-\Id\cc{i}{1}\cc{j}{2}\cc{k}{3}\right) m_j^2+\left(-\Id\cc{i}{1}\cc{j}{3}\cc{k}{2}-\Id\cc{i}{2}\cc{j}{2}\cc{k}{2}-\Id\cc{i}{3}\cc{j}{1}\cc{k}{2}+\I\cc{i}{1}\cc{j}{2}\cc{k}{2}+\I\cc{i}{2}\cc{j}{1}\cc{k}{2}\right) m_k^2 \, , \nonumber \\
 f_{12,a}^{ij} &=& \frac{1}{3} \left(2 \Iq\cc{i}{2}\cc{j}{5}-3 \Id\cc{i}{2}\cc{j}{4}+\Id\cc{i}{3}\cc{j}{3}+\Id\cc{i}{4}\cc{j}{2}+\I\cc{i}{2}\cc{j}{3}-\I\cc{i}{3}\cc{j}{2}\right) m_i^2 \, , \nonumber \\
 f_{12,b}^{ij} &=&   f_{12,a}^{ij} \, ,  \nonumber \\
 f_{12,c}^{ij} &=& \frac{1}{3} \left(-3 \Id\cc{i}{2}\cc{j}{4}+\I\cc{i}{2}\cc{j}{3}+2\Iq\cc{i}{2}\cc{j}{5}-2\Id\cc{i}{3}\cc{j}{3}-2\Id\cc{i}{4}\cc{j}{2}+2\I\cc{i}{3}\cc{j}{2}\right) m_i^2 \, , \nonumber \\
 f_{13}^{ijk} &=& \frac{1}{2} \Big(\left(\I\cc{i}{3}\cc{j}{1}\cc{k}{1}-\Id\cc{i}{4}\cc{j}{1}\cc{k}{1}\right) m_i^2+\left(-\Id\cc{i}{1}\cc{j}{4}\cc{k}{1}-\Id\cc{i}{2}\cc{j}{3}\cc{k}{1}-\Id\cc{i}{3}\cc{j}{2}\cc{k}{1}+\I\cc{i}{1}\cc{j}{3}\cc{k}{1}+\I\cc{i}{2}\cc{j}{2}\cc{k}{1}\right) m_j^2 \nonumber \\
 && \qquad +\left(-\Id\cc{i}{1}\cc{j}{1}\cc{k}{4}-\Id\cc{i}{1}\cc{j}{2}\cc{k}{3}-\Id\cc{i}{1}\cc{j}{3}\cc{k}{2}-\Id\cc{i}{2}\cc{j}{1}\cc{k}{3}-\Id\cc{i}{2}\cc{j}{2}\cc{k}{2}-\Id\cc{i}{3}\cc{j}{1}\cc{k}{2}+\I\cc{i}{1}\cc{j}{1}\cc{k}{3}+\I\cc{i}{1}\cc{j}{2}\cc{k}{2}+\I\cc{i}{2}\cc{j}{1}\cc{k}{2}\right) m_k^2\Big) \, , \nonumber \\
 f_{14}^{ijk} &=& \left(\Id\cc{i}{1}\cc{j}{3}\cc{k}{2}+2 \Id\cc{i}{1}\cc{j}{4}\cc{k}{1}+\Id\cc{i}{2}\cc{j}{3}\cc{k}{1}-2 \I\cc{i}{1}\cc{j}{3}\cc{k}{1}\right) m_j^2 \, , \nonumber \\
 f_{15a}^{ijk} &=&  \frac{1}{3} \left(-\Id\cc{i}{3}\cc{j}{1}\cc{k}{2}-2 \Id\cc{i}{4}\cc{j}{1}\cc{k}{1}+2 \I\cc{i}{3}\cc{j}{1}\cc{k}{1}\right) m_i^2 \nonumber \\
 && + \frac{1}{3}\left(-\Id\cc{i}{1}\cc{j}{3}\cc{k}{2}-2 \Id\cc{i}{1}\cc{j}{4}\cc{k}{1}-\Id\cc{i}{2}\cc{j}{2}\cc{k}{2}+2 \I\cc{i}{1}\cc{j}{3}\cc{k}{1}+2 \I\cc{i}{2}\cc{j}{2}\cc{k}{1}-2 \Id\cc{i}{2}\cc{j}{3}\cc{k}{1}-2\Id\cc{i}{3}\cc{j}{2}\cc{k}{1}\right) m_j^2 \nonumber \\
 && +\frac{1}{3}\left(2 \Id\cc{i}{1}\cc{j}{1}\cc{k}{4}+\Id\cc{i}{1}\cc{j}{2}\cc{k}{3}+\Id\cc{i}{2}\cc{j}{1}\cc{k}{3}-2 \I\cc{i}{1}\cc{j}{1}\cc{k}{3}\right) m_k^2  \, , \nonumber \\
  f_{15b}^{ijk} &=& \frac{1}{3} \left(\Id\cc{i}{3}\cc{j}{1}\cc{k}{2}+\Id\cc{i}{3}\cc{j}{2}\cc{k}{1}+2 \Id\cc{i}{4}\cc{j}{1}\cc{k}{1}-2 \I\cc{i}{3}\cc{j}{1}\cc{k}{1}\right) m_i^2 \nonumber \\
  && + \frac{1}{3} \left(-2 \Id\cc{i}{1}\cc{j}{4}\cc{k}{1}-\Id\cc{i}{2}\cc{j}{3}\cc{k}{1}+2 \I\cc{i}{1}\cc{j}{3}\cc{k}{1}\right) m_j^2 \nonumber \\
 && +\frac{1}{3} \left(-2 \Id\cc{i}{1}\cc{j}{1}\cc{k}{4}-2 \Id\cc{i}{1}\cc{j}{2}\cc{k}{3}-2 \Id\cc{i}{1}\cc{j}{3}\cc{k}{2}-\Id\cc{i}{2}\cc{j}{1}\cc{k}{3}-\Id\cc{i}{2}\cc{j}{2}\cc{k}{2}+2 \I\cc{i}{1}\cc{j}{1}\cc{k}{3}+2 \I\cc{i}{1}\cc{j}{2}\cc{k}{2} \right) m_k^2 \, , \nonumber \\
 f_{16}^{ijklm} &=& \I\cc{i}{2}\cc{j}{1}\cc{k}{1}\cc{l}{1}\cc{m}{1} m_{i}^{2} \, , \nonumber \\
 f_{17}^{ijkl} &=& \left(-\Id\cc{i}{2}\cc{j}{1}\cc{k}{1}\cc{l}{3}+\I\cc{i}{2}\cc{j}{1}\cc{k}{1}\cc{l}{2}\right) m_i^2 +\left(\I\cc{i}{1}\cc{j}{2}\cc{k}{1}\cc{l}{2}-\Id\cc{i}{1}\cc{j}{2}\cc{k}{1}\cc{l}{3}\right) m_j^2 +\left(\I\cc{i}{1}\cc{j}{1}\cc{k}{2}\cc{l}{2}-\Id\cc{i}{1}\cc{j}{1}\cc{k}{2}\cc{l}{3}\right) m_k^2 \nonumber \\
 && +\big(-\Id\cc{i}{1}\cc{j}{1}\cc{k}{3}\cc{l}{2}-\Id\cc{i}{1}\cc{j}{2}\cc{k}{2}\cc{l}{2}-\Id\cc{i}{1}\cc{j}{3}\cc{k}{1}\cc{l}{2}-\Id\cc{i}{2}\cc{j}{1}\cc{k}{2}\cc{l}{2}-\Id\cc{i}{2}\cc{j}{2}\cc{k}{1}\cc{l}{2}-\Id\cc{i}{3}\cc{j}{1}\cc{k}{1}\cc{l}{2} \nonumber \\
 && +\I\cc{i}{1}\cc{j}{1}\cc{k}{2}\cc{l}{2}+\I\cc{i}{1}\cc{j}{2}\cc{k}{1}\cc{l}{2}+\I\cc{i}{2}\cc{j}{1}\cc{k}{1}\cc{l}{2}\big) m_l^2
 \, , \nonumber \\
 f_{18}^{ijkl} &=& \left(-\Id\cc{i}{2}\cc{j}{1}\cc{k}{1}\cc{l}{3}-\Id\cc{i}{2}\cc{j}{1}\cc{k}{2}\cc{l}{2}-\Id\cc{i}{2}\cc{j}{1}\cc{k}{3}\cc{l}{1}+\I\cc{i}{2}\cc{j}{1}\cc{k}{1}\cc{l}{2}+\I\cc{i}{2}\cc{j}{1}\cc{k}{2}\cc{l}{1}\right) m_i^2 \nonumber \\
 && +\left(-\Id\cc{i}{1}\cc{j}{3}\cc{k}{1}\cc{l}{2}-\Id\cc{i}{2}\cc{j}{2}\cc{k}{1}\cc{l}{2}-\Id\cc{i}{3}\cc{j}{1}\cc{k}{1}\cc{l}{2}+\I\cc{i}{1}\cc{j}{2}\cc{k}{1}\cc{l}{2}+\I\cc{i}{2}\cc{j}{1}\cc{k}{1}\cc{l}{2}\right) m_l^2
 \, , \nonumber \\
 f_{19}^{ijklmn} &=& \I\cc{i}{2}\cc{j}{1}\cc{k}{1}\cc{l}{1}\cc{m}{\hspace{0.4mm} 1}\cc{n}{1} m_{i}^{2} \, . 
\label{eq:fexpressions}
\end{eqnarray}
}
These $f_{N}^{ijk\cdots}$ coefficients are elementary building blocks, in the sense that any Wilson coefficient of a given theory
is constituted by some combination of these universal coefficients. 

The expressions (\ref{eq:fexpressions}) are quite compact. They contain all the dependences on the masses $m_i$ of the heavy particles 
$\Phi_i$ that have been integrated out, and simplify considerably when these masses are degenerate, $m_i = m$: 
\begin{eqnarray}
 f_{5} =  -\frac{i}{(4\pi)^{2} 60 m^2} , \quad  & f_{11} =  \frac{i}{(4\pi)^{2} 12 m^4} , \quad & f_{15a} = \frac{i}{(4\pi)^{2} 60 m^4} \, ,\nonumber \\
 f_{6} =  -\frac{i}{(4\pi)^{2}90 m^2} , \quad & f_{12,a} =  0 , \quad & f_{15b}= \frac{i}{(4\pi)^{2} 60 m^4} \, , \nonumber \\ 
 f_{7} =  -\frac{i}{(4\pi)^{2}12 m^2} , \quad & f_{12,b} =  0 , \quad & f_{16} = -\frac{i}{(4\pi)^2 60 m^6} \, , \nonumber \\ 
 f_{8} =  -\frac{i}{(4\pi)^{2}6 m^2} , \quad & f_{12,c} =  \frac{i}{(4\pi)^{2} 120 m^4} , \quad & f_{17} = -\frac{i}{(4\pi)^2 20 m^6} \, , \nonumber \\ 
 f_{9} =  -\frac{i}{(4\pi)^{2}12 m^2} , \quad & f_{13} =  \frac{i}{(4\pi)^{2} 24 m^4}  , \quad & f_{18} = -\frac{i}{(4\pi)^2 30 m^6} \nonumber \\ 
 f_{10} =  \frac{i}{(4\pi)^{2}24 m^4} , \quad & f_{14} = \frac{-i}{(4\pi)^{2} 60 m^4} , \quad & f_{19} =  \frac{i}{(4\pi)^2 120 m^8}  \, .
 \label{eq:degf}
\end{eqnarray}

We see that the operators associated with $f_{12,a/b}$, i.e., the operators
$\left[P_{\mu},[P_{\nu},U]\right]\left[P_{\mu},[P_{\nu},U]\right]$ and $\left[P_{\mu},[P_{\nu},U]\right]\left[P_{\nu},[P_{\mu},U]\right]$, 
do not contribute in the degenerate case. We have checked that the universal coefficients in the degenerate limit, together with Eq.~(\ref{eq:universallagrangian}), 
recover the results of Eq.~(2.3) of Ref.~\cite{HLM}. In the general non-degenerate case the mass dependences of the universal coefficients can be seen in 
Appendix~\ref{app:fexplicit}, where we write out explicitly Eq.~(\ref{eq:fexpressions}) with the integrals evaluated.

The first four coefficients $f_{1,2,3,4}$ in (\ref{eq:fexpressions}) exhibit UV divergences, and involve $\mu$, the usual scale of renormalization, and dimensions $d$ as in $f_3$,  
which has been introduced through dimensional regularization using the $\overline{\text{MS}}$ scheme.
All these terms are already present in the Standard Model, and so can be absorbed in redefinitions of the parameters of the Standard Model. 
The first coefficient, $f_1$, which is a simple constant of mass dimension four, is a renormalization of the vacuum energy
that can be absorbed as a constant term in the Standard Model Higgs potential. Depending on the form of the $U$-matrix, 
the coefficients $f_2$ and $f_4$, of mass dimension two and zero respectively, may renormalize some other Standard Model couplings. 
The third coefficient, $f_3$, of mass dimension zero, is a new contribution to the gauge kinetic terms that must be removed by a wave-function
renormalization. Usually, this induces a renormalization of the gauge couplings so as to keep the canonical form of the covariant derivatives. 
Each divergence introduces a relation between the bare parameters of the Standard Model and the corresponding bare parameters of the EFT.
The lowest-dimensional operators generated by the last nine lines of (\ref{eq:universallagrangian}) have dimension of at least 6, since they
have coefficients $f_{N}^{ijk\cdots}$ of mass dimension $-4$ for $f_{5}$ to $f_{9}$, $-6$ for the coefficients $f_{10}$ to $f_{15}$, 
$-8$ for the coefficients $f_{16}$ to $f_{18}$ and $-10$ for the coefficient $f_{19}$. These coefficients, $f_{5}$ to $f_{19}$, are free of any divergences.

In Appendix \ref{app:stop} we work out a concrete example, by applying our universal one-loop effective Lagrangian to integrate out the third-generation squark doublet and stop singlet in the MSSM to obtain the Wilson coefficients in the SM EFT. This provides some insight into how each term in (\ref{eq:universallagrangian}), and their universal coefficients (\ref{eq:fexpressions}), may combine to give the final Wilson coefficient of EFT operators in a realistic example. As a cross-check we also compare the MSSM coefficients that we obtain using our method with previous results in the literature.

We insist one more time on the originality and power of our approach by contrasting with two other ways for obtaining Wilson coefficients at one-loop order: 
\begin{enumerate} 
	\item Starting from the Lagrangian for the UV theory, derive the Feynman rules, compute Feynman diagrams involving loop integrals for the relevant observable, and do the same in the EFT. Compare the two calculations to extract the Wilson coefficient. Repeat for each coefficient. 
	\item Starting from the Lagrangian for the UV theory, rewrite the path integral involving the quadratic term in the heavy field as a logarithm, expand the logarithm using a method such as the CDE, then evaluate the nested series of derivatives acting on momenta between the various non-commuting matrices which gets increasingly complicated and tedious to compute. Finally, perform the numerous integrals over momenta for each of the many terms involving different Lorentz index contractions of covariant derivatives with momentum tensors, before finally combining them into the desired operators to get their Wilson coefficients. 
\newcounter{enumTemp}
\setcounter{enumTemp}{\theenumi}
\end{enumerate}
With the universality property of the path-integral method noticed by HLM for the degenerate case~\cite{HLM}, and extended here to the general case, we may instead summarise the new way to compute one-loop Wilson coefficients as follows: 
\begin{enumerate} 
\setcounter{enumi}{\theenumTemp}
\item Starting from the Lagrangian for the UV theory, write it in the form of Eq.~(\ref{eq:lagrangianUV}) to extract the $U$ matrix of light fields, the covariant derivative $P_\mu$, and the field strength matrix $G^\prime_{\mu\nu}$. Input this into Eq.~(\ref{eq:universallagrangian}) and sum over the matrix indices. Arrange the result into the desired operator basis to obtain the Wilson coefficients. 
\end{enumerate}
This technique bypasses the lengthy initial steps of the previous two methods and avoids redoing everything from beginning to end for each model. Clearly, the universal coefficients could not have been computed by the usual Feynman diagram method that requires explicit Feynman rules for a particular model~\footnote{However, the Feynman diagram method still holds an advantage when
matching with higher precision beyond the one-loop level. We note that while the effects of mixed light and heavy fields in loops is typically performed by further matching using Feynman diagrams~\cite{bilenkysantamaria}, it has been shown that functional methods may also be used to compute these~\cite{mixedlightheavy}. }, and though the CDE method is an elegant way of obtaining directly the EFT from a UV theory, it is no longer necessary for the purpose of matching at one-loop.

\section{The Covariant Derivative Expansion}
\label{sec:CDE}

As discussed in the previous Section, the reader who wishes to compute one-loop Wilson coefficients for operators of dimension up to six can start directly with (\ref{eq:universallagrangian}) and need not worry about the details of the CDE method that we used to derive the universal one-loop effective Lagrangian. However, there are many other cases where one may wish to use the path integral, so we briefly summarise the CDE here. This was first introduced in the 1980s by Gaillard~\cite{Gaillard} and Cheyette~\cite{Cheyette}. We refer to the extensive review in Ref.~\cite{HLM} for a clear and detailed description.

\subsection{The CDE Method for Integrating out Fields}

Starting from an action $S[\phi,\Phi]$ for the UV theory, where $\phi$ and $\Phi$ represent the light and heavy fields respectively, we may expand around the minimum and evaluate the path integral for the heavy fields. For scalar fields the effective action can then be written as
\begin{align*}
e^{iS_\text{eff}[\phi]} &= \int [D\Phi] e^{iS[\phi,\Phi]} \\
&= \int [D\eta] e^{i\left(S[\phi,\Phi_c] + \frac{1}{2}\left.\frac{\delta^2 S}{\delta \Phi^2}\right|_{\Phi=\Phi_c}\eta^2 + \mathcal{O}(\eta^3)\right)} \\
&\approx e^{iS[\phi,\Phi_c]}\left[\text{det}\left(\left.-\frac{\delta^2 S}{\delta\Phi^2}\right|_{\Phi=\Phi_c}\right)\right]^{-\frac{1}{2}}	\\
&\approx e^{iS[\phi,\Phi_c] - \frac{1}{2}\text{Tr ln}\left(-\left.\frac{\delta^2 S}{\delta \Phi}\right|_{\Phi=\Phi_c}\right)} \, ,
\end{align*}
where $\Phi_c$ is defined as $\left.\frac{\delta S}{\delta \Phi}\right|_{\Phi = \Phi_c} = 0$. This can be applied to bosons or fermions, and the result is in general a one-loop part of the effective action of the form 
\begin{equation*}
S_{\text{1-loop}}^{\text{eff}} = i c_s \text{Tr ln}\left( -P^2 + M^2 + U \right)	\, . 
\label{Seff}
\end{equation*}
As discussed above, the constant $c_{s}$ depends on whether one is integrating out real scalars, complex scalars, 
Majorana or Dirac fermions, gauge bosons or Fadeev-Popov ghosts, for which it takes the values $=1/2,1,-1/2,1/2$ and $-1$ respectively,
and $P_\mu \equiv iD_\mu$ is the covariant derivative. The form of $U$ obtained from the original UV Lagrangian also depends on the type of field(s) being integrated out~\cite{HLM}. 

After the trace over space-time is evaluated by inserting a complete set of spatial and momentum eigenstates, we are left with a trace ``tr'' over internal indices only, 
\begin{equation*}
S_{\text{1-loop}}^{\text{eff}} = i c_s\int d^4x \int\frac{d^4q}{(2\pi)^4} \text{tr ln}\left( -(P_\mu - q_\mu)^2 + M^2 + U \right)	\, . 
\end{equation*}
Before expanding the logarithm, it is convenient to shift the momentum in the integral using the covariant derivative by inserting factors of $e^{\pm P_\mu\partial/\partial q_\mu}$: 
\begin{equation*}
\mathcal{L}^\text{eff}_\text{1-loop} = i c_{s} \int \frac{d^4q}{(2\pi)^4} \text{tr} \ln[e^{P_\mu\partial/\partial q_\mu}(-(P_\mu-q_\mu)^2 + M^2 + U) e^{-P_\mu\partial/\partial q_\mu} ] \, .
\end{equation*}
This choice ensures an expansion that involves only manifestly 
gauge-invariant pieces throughout. The result is a series involving gauge field strengths, 
covariant derivatives and the SM fields encoded in the matrix $U(x)$:
\begin{equation}
\mathcal{L}^\text{eff}_\text{1-loop} = i c_{s} \int \frac{d^4q}{(2\pi)^4} \text{tr} \ln[-(\tilde{G}_{\nu\mu}\partial/\partial q_\mu + q_\mu)^2 + M^2 + \tilde{U}]	\, ,
\label{Leff_ln}
\end{equation}
where
\begin{align*}
\tilde{G}_{\nu\mu} &\equiv \sum_{n=0}^{\infty} \frac{n+1}{(n+2)!}[P_{\alpha_1},[...[P_{\alpha_n},G^\prime_{\nu\mu}]]]\frac{\partial^n}{\partial q_{\alpha_1} ... q_{\alpha_n}} 	\, , \\
\tilde{U} &=\sum_{n=0}^{\infty} \frac{1}{n!}[P_{\alpha_1},[...[P_{\alpha_n},U]]]	\frac{\partial^n}{\partial q_{\alpha_1} ... q_{\alpha_n}} \, ,
\end{align*}
where we define $G^\prime_{\nu\mu}$ as the field strength given by 
$[P_\nu,P_\mu] = -G^\prime_{\nu\mu}$. 

We may now expand the logarithm in (\ref{Leff_ln}) to obtain an explicit sum.
When all the heavy fields are degenerate in mass $m$, 
one can easily expand the action by integrating once its derivative 
with respect to the common mass scale $m$,
\begin{align} 
 \mathcal{L}^\text{eff}_\text{1-loop}
=  -i c_{s} \int \frac{d^4q}{(2\pi)^4} \int dm^2  \text{tr} \left[  \frac{1}{  (q^2-m^2+\{ q_\mu,\tilde{G}_{\nu \mu}\}\frac{\partial}{\partial q_\nu}+\tilde{G}_{\sigma\mu}\tilde{G}^\sigma {}_\nu \frac{\partial}{\partial q_\mu} \frac{\partial}{\partial q_\nu} -\tilde{U}) } \right]\mathds{1} \, .
\label{expanding_action_1}
\end{align}
The integral over mass is not of physical significance and may be regarded as integrating over a spurious parameter, as we shall see in the non-degenerate case. An alternative is to expand the logarithm directly using the Baker-Campbell-Hausdorff formula as in Ref.~\cite{Cheyette}. Here we choose to follow Ref.~\cite{HLM} for convenience of comparison when we generalise to the non-degenerate case. 

Defining $\Delta \equiv 1/(q^2-m^2)$, which is the free propagator of the
massive field that is integrated out, one can Taylor expand to obtain the following effective Lagrangian:
\begin{eqnarray}
\mathcal{L}^\text{eff}_\text{1-loop} &=&  -i c_{s} \int \frac{d^4q}{(2\pi)^4} \int dm^2  \text{tr} \Big\{  
\Delta 
-\Delta \left( \{ q_\mu,\tilde{G}_{\nu \mu}\} \frac{\partial}{\partial q_\nu}+\tilde{G}_{\sigma\mu}\tilde{G}^\sigma {}_\nu \frac{\partial}{\partial q_\mu} \frac{\partial}{\partial q_\nu} -\tilde{U} \right) \Delta \nonumber \\
&& \hspace{-1.5cm} + \Delta \left( \{ q_\mu,\tilde{G}_{\nu \mu}\}\frac{\partial}{\partial q_\nu}+\tilde{G}_{\sigma\mu}\tilde{G}^\sigma {}_\nu \frac{\partial}{\partial q_\mu} \frac{\partial}{\partial q_\nu} -\tilde{U} \right) \Delta \left( \{ q_\mu,\tilde{G}_{\nu \mu}\}\frac{\partial}{\partial q_\nu}+\tilde{G}_{\sigma\mu}\tilde{G}^\sigma {}_\nu \frac{\partial}{\partial q_\mu} \frac{\partial}{\partial q_\nu} -\tilde{U} \right) \Delta \nonumber \\
&& + ... \Big\} \, .
\label{expanding_action_2}
\end{eqnarray}
Since we are considering, at this stage, only the special case where the heavy fields $\Phi$
are degenerate in mass, the mass-squared matrix is proportional to the identity, $M^{2} = m^2\identity$. 
Therefore, $\Delta$ commutes with the matrices $\tilde{U}$ and $\tilde{G}$, and (\ref{expanding_action_2})
may simply be rewritten as
 \begin{align} 
\mathcal{L}^\text{eff}_\text{1-loop} = 
-i c_{s} \int \frac{d^4q}{(2\pi)^4} \int dm^2 \text{tr} \left[ \sum_{n=0}^{\infty} \left( \{ q_\mu,\tilde{G}_{\nu \mu}\}\frac{\partial}{\partial q_\nu}+\tilde{G}_{\sigma\mu}\tilde{G}^\sigma {}_\nu \frac{\partial}{\partial q_\mu} \frac{\partial}{\partial q_\nu} -\tilde{U} \right)^n  (-\Delta)^{n+1}  \right] \mathds{1} \, .
\label{expanding_action_3}
\end{align}
In this case, the integration over momentum in
each term of the expansion factorizes out of the trace, yielding the degenerate universal coefficients of the higher-dimensional operators that originate from the various combinations of $U$ and $G^\prime_{\mu\nu}$ in the series of summations. 

This method is completely general, assuming only the mass degeneracy that we generalize in the next sub-section, but for the purpose of computing the universal coefficients we restrict our attention to the terms in (\ref{expanding_action_3}) up to $n=6$ that yield all possible combinations of light fields,  encapsulated in $U$ and $G^\prime_{\mu\nu}$, for operators of dimension up to six.
Terms of higher orders in $n$ will be responsible for higher-dimensional operators. 

\subsection{Integrating out Non-Degenerate Fields}

The main objective of this work is to extend the universal expansion that we have just described in the degenerate mass situation 
to the general situation where fields are non-degenerate in mass. 
In the non-degenerate case, we can no longer replace the logarithm in (\ref{Leff_ln})
by a single additional integration over mass as was done in deriving (\ref{expanding_action_1}). 

As mentioned before, one may use the Baker-Campbell-Hausdorf formula and expand directly
the logarithm~\cite{Cheyette,DEQY1} to evaluate the expansion terms of (\ref{Leff_ln}) in the presence of non-commuting matrices.

Alternatively, in the following we introduce an auxiliary parameter $\xi$ in order to evaluate the expansion terms. The parameter $\xi$ multiplies the diagonal mass matrix $M$, 
\beq
 M= \xi \cdot {\rm Diag}(m_{i}) \, ,
\eeq
that we can now differentiate and integrate over, before setting $\xi$ to 1 at the end of the computation when the integration has been performed.
Then, in the non-degenerate case, (\ref{Leff_ln}) is replaced by
\begin{eqnarray} 
&& \hspace {-0.5cm}
\mathcal{L}^\text{eff}_\text{1-loop}
=  -i c_{s} \int \frac{d^4q}{(2\pi)^4} \int d\xi  \text{tr} \left[  \frac{1}{  (\Delta_{\xi}^{-1}+\{ q_\mu,\tilde{G}_{\nu \mu}\}\frac{\partial}{\partial q_\nu}+\tilde{G}_{\sigma\mu}\tilde{G}^\sigma {}_\nu \frac{\partial}{\partial q_\mu} \frac{\partial}{\partial q_\nu} -\tilde{U}) } M^{2} \right] \, ,
\label{expanding_action_4}
\end{eqnarray}
where we have adapted our previous notation for the free propagator, introducing $\Delta_{\xi} \equiv 1/(q^2-\xi M^2)$.
The Taylor expansion then gives
\begin{align}
\mathcal{L}^\text{eff}_\text{1-loop} = -i c_{s} \int \frac{d^4q}{(2\pi)^4} \int d\xi \text{tr}\left\{ \sum_{n=0}^\infty \left[-\Delta_\xi(\{ q_\mu,\tilde{G}_{\nu \mu}\}\frac{\partial}{\partial q_\nu}+\tilde{G}_{\sigma\mu}\tilde{G}^\sigma {}_\nu \frac{\partial}{\partial q_\mu} \frac{\partial}{\partial q_\nu} -\tilde{U})\right]^n\Delta_\xi M^2 \right\}	\, .
\label{expanding_action_5}
\end{align}
Since we are now in the general case where the field components are non-degenerate in mass, 
{\it a priori} the diagonal mass-squared matrix $M^{2}$ and the propagator $\Delta_{\xi}$ 
do not commute with the matrices $\tilde{U}$ and $\tilde{G}$. 
As a result, the integrals in (\ref{expanding_action_5}) are quite involved to evaluate.

In order to deal with these non-commuting objects, and recover some of the simplicity
of the previous degenerate case, we decompose explicitly the traces of the matrix products
and work with matrix elements that commute in general. This then allows us to factor out the momentum-dependent integrals. 
In the degenerate case, the cyclicity of the trace is commonly used in order to rewrite some operators. 
However, care must be taken in the non-degenerate case, where we shift the indices of the various matrix elements.

It is important to note that the mass matrix $M$ is not necessarily the physical mass matrix, e.g., for our example in appendix D with the MSSM stops, the mass matrix $M$ contains the supersymmetry-breaking masses, which are related through diagonalization to the physical stop masses. However, it could equally well be the physical mass matrix, if one diagonalises the mass matrix $M$ before integrating out the heavy fields. We recall that it is essential for many cases of physical interest to be able to relax the commutativity condition between the matrices $U$ and $M$, for example when different multiplets are connected through off-diagonal terms in the quadratic interaction matrix, or if one writes the Lagrangian in the physical mass basis after EWSB before integrating out at one-loop then there are generally non-degeneracies within the multiplets themselves.

\section{Discussion}
\label{sec:conclusion}

We stress once more that the results given above are universal, in the sense that the dependences on non-degenerate masses in loop integrals factor out of generic operator structures in the one-loop effective action, so that our results apply whenever
any multiplets of massive particles are integrated out. The obvious application is
to massive electroweak doublets and singlets as occur in many phenomenological extensions of the SM, which is exemplified in Appendix~\ref{app:stop},
but our results have more general applicability. We emphasize that the universal one-loop effective action is equally
applicable to loops of massive bosons, e.g., the sfermions of supersymmetric models, and to
loops of massive fermions, e.g., vector-like fermions. Furthermore, it is not restricted to the
effective action for external bosons, but can also be used for external fermions. In light of its generality, we suggest that future calculations of 
one-loop Wilson coefficients for operators of dimenson up to six skip the usual Feynman diagram or path-integral methods and proceed directly to our Eq.~(\ref{eq:universallagrangian}) as the starting point.

It is instructive to review the connection of our work with the SM EFT approach. One obvious advantage of using the universal one-loop effective action is the systematic way in which all effective operators may be obtained once the form of the covariant derivative, $P_\mu$, and the matrices, $U$ and $G^\prime_{\mu\nu}$, are specified~\footnote{Writing the UV Lagrangian in the appropriate form for $U$ is trivial in the scalar case, but requires some manipulations when integrating out vector bosons and fermions, see Appendix~\ref{app:fermions} and Ref.~\cite{HLM}.}, where $G^\prime_{\mu\nu}$ is related to the usual field strength, while $U$ encodes the dependence on the couplings to light fields for the specific form of the UV model. For example, if one is interested in obtaining the experimental constraints from the popular $S$ and $T$~\cite{PeskinTakeuchi} (or equivalently $\epsilon_{1,2}$~\cite{AltarelliBarbieri})
parameters for a particular model, not only is this the easiest way to calculate one-loop contributions to the relevant Wilson coefficients, 
assuming decoupled new physics, but it can also provide complementary coefficients for other observables with little extra effort. 
Bounds can then be set on this model from other SM measurements or, if the error bars are larger, to set target precisions in these other measurements.  

The `killer application' of this method is to integrating out particles with non-degenerate masses, 
which was a limitation of the original universal one-loop effective action of Ref.~\cite{HLM} that applied to degenerate masses. 
There are many such scenarios with particles occurring in unbroken SM gauge multiplets, each with different masses. 
In particular, they may have non-degenerate masses within a given multiplet. 
If the masses of the degrees of freedom that are integrated out are not extremely high, electroweak symmetry
breaking will, in general cause significant splittings within electroweak multiplets. Our one-loop effective
action can explicitly include all the effects of such possible non-degeneracies. In the SM EFT
approach, if integrating out a degenerate multiplet would give an operator of some dimension
$n$, non-degeneracies within this multiplet would in general yield a series of operators with
dimensions $\ge n$ corresponding to the lowest-dimensional operator supplemented by external
Higgs fields with vacuum expectation values $v$. Therefore, truncating the SM EFT operator expansion
at some fixed dimension, e.g., $n = 6$ as is often done in phenomenological analyses, does not in general
yield a complete description of the low-energy physics due to integrating out non-degenerate electroweak multiplets~\footnote{See Refs.~\cite{craigetal, gorbahnetal, DEQY1, brehmeretal} for some studies of differences between EFT and full calculations.}. Of course, this discrepancy becomes less important when the heavy particles have masses
that are much larger than the Higgs vacuum expectation value $v$.

In practice, the present constraints on the coefficients of dimension-6 operators are
sufficiently weak that the operators could be generated by loops of particles that are {\it not}
much heavier than $v$~\footnote{This situation may well change in the future if LHC and other
data increase significantly the current lower limits on particles beyond the SM.}. 
In this case, although the dimension-6 SM EFT constraints seem appealingly
universal, they may not be applicable to specific models in which electroweak symmetry breaking
could induce non-negligible non-degeneracies. A case in point is supersymmetry:
both the direct and indirect constraints on stop masses, for example, are consistent with
the lighter stop mass being ${\cal O}(v)$, whereas the heavier stop might be much heavier.
We considered in~\cite{DEQY1} constraints on stop masses as obtained using
the dimension-6 SM EFT and exact one-loop calculations. Whilst the constraints were broadly
comparable, there were significant differences for stop masses below $\sim 500$ GeV~\cite{craigetal, DEQY1}, and the SM EFT approach is then of limited use for
analyzing light-stop scenarios. The same would be true for other light-sparticle scenarios, e.g., 
models with light neutralinos and charginos.

The universal one-loop effective action set out here could be used to analyze
these and other scenarios with relatively light BSM particles, and we are preparing a
more complete analysis of the one-loop constraints on supersymmetric models,
building upon the analysis in~\cite{DEQY1}, which considered just the most
important constraints on light stop and sbottom squarks. A first step towards this is presented in Appendix~\ref{app:stop}, which lists the Wilson coefficients 
that we obtain by integrating out non-degenerate squarks, which provides non-trivial cross-checks of the 
universal coefficients of Eq.~(\ref{eq:universallagrangian}). Other possible candidates for
studies using the universal one-loop effective action include scenarios with extra 
vector-like fermions~\cite{RHvectorlike, vectorlikesurvey}, and composite resonances or massive gauge bosons from new strong sector and extra-dimensional models~\cite{compositealternatives}, for example. It will also be interesting to explore further the connection between the SM EFT and 
measurements of $B$-meson decays~\cite{Bphysics}.  

The interplay between direct and indirect searches will only get increasingly important as the experimental sensitivity in many measurements reaches the multi-TeV scale
both at the LHC~\cite{goldenratio} and, eventually, at future lepton colliders~\cite{EYfuture}. Even if  BSM resonances are found in Run 2 of the LHC or after~\cite{100TeV}, it is likely that this will only yield partial information. 
Only by complementing such a new discovery with more precise indirect measurements can we gain a fuller picture of the new sector. The SM EFT
and the universal one-loop effective action can play a valuable role in such analyses. 

In view of its general applicability, we are making available on request a
{\tt Mathematica} notebook with explicit expressions for the universal coefficients and their integrals.

\section*{Acknowledgements}

We thank Brian Henning and Ran Huo for valuable communications, and TY thanks Dave Sutherland for helpful discussions. JQ and TY are grateful to CERN for hospitality while this work was carried out. We also thank Herm\`es B\'elusca-Ma\"{i}to and Zhengkang Zhang  for pointing out typos in the preprint version 1. 
The work of AD was supported by the STFC Grant ST/J002798/1. 
The work of JE was supported partly by the London Centre
for Terauniverse Studies (LCTS), using funding from the European Research Council via the Advanced Investigator
Grant 26732, and partly by the STFC Grants ST/J002798/1 and ST/L000326/1.
The work of JQ was supported by the STFC Grant ST/L000326/1. 
The work of TY was supported initially by a
Graduate Teaching Assistantship from King's College London
and now by a Junior Research Fellowship from Gonville and Caius College, Cambridge.

\begin{appendices}
\section{Computation of the Universal Coefficients $f_{N}^{ijk...}$}
\label{app:fcomputation}

We give here more details concerning the evaluation of the universal coefficients, $f_{N}^{ijk...}$, appearing in (\ref{eq:universallagrangian}). 
We try to keep the notation of \cite{HLM} as much as possible, in order to clarify the relation between the two results. 

The CDE allowed us to derive (\ref{expanding_action_5}) that we can conveniently re-write as:
\beq
\mathcal{L}^\text{eff}_\text{1-loop} = -i c_{s} \text{tr} \left\{ M^{2}  \sum_{n=0}^\infty  {\cal I}_{n} \right\}
\eeq
with
\begin{align}
{\cal I}_{n} =   \int \frac{d^4q}{(2\pi)^4}  \int d\xi \left[-\Delta_\xi(\{ q_\mu,\tilde{G}_{\nu \mu}\}\frac{\partial}{\partial q_\nu}+\tilde{G}_{\sigma\mu}\tilde{G}^\sigma {}_\nu \frac{\partial}{\partial q_\mu} \frac{\partial}{\partial q_\nu} -\tilde{U})\right]^n \Delta_\xi \, .
\end{align}
When performing the calculation it is useful to distinguish the ${\cal I}_{n}$ integrals into those involving only $\tilde{G}$, called $J_{n}$, 
or involving only $\tilde{U}$, called $K_{n}$: 
\begin{align}
J_{n} &=  \int \frac{d^4q}{(2\pi)^4}  \int d\xi \left[-\Delta_\xi(\{ q_\mu,\tilde{G}_{\nu \mu}\}\frac{\partial}{\partial q_\nu}+\tilde{G}_{\sigma\mu}\tilde{G}^\sigma {}_\nu \frac{\partial}{\partial q_\mu} \frac{\partial}{\partial q_\nu})\right]^n \Delta_\xi \, , \nonumber \\
K_{n} &=  \int \frac{d^4q}{(2\pi)^4}  \int d\xi \left[\Delta_\xi \tilde{U} \right]^n \Delta_\xi \, ,
\end{align}
with integrals involving mixed $\tilde{U}$ and $\tilde{G}$ terms are called  $L_{n}$. 
Each integral contains one or several operators of dimension up to six. The mass-dependent parts (equivalently the momentum integral terms) 
can be factorized and constitute what we denote as the $f_{N}^{ijk...}$ coefficients. 

\begin{table}[h!]
\begin{center}
\begin{tabular}{|l|c|r|}
  \hline
  Integrals & Coefficients \\
  \hline
  $J_{1}$ & $f_{3}$, $f_{5}$, $f_{6}$ \\
  $J_{2}$ & $f_{5}$ \\  
   \hline
  $K_{1}$ & $f_{2}$ \\ 
  $K_{2}$ & $f_{4}$, $f_{7}$, $f_{12a}$, $f_{12b}$, $f_{12c}$ \\ 
  $K_{3}$ & $f_{8}$, $f_{11}$ \\ 
  $K_{4}$ & $f_{10}$, $f_{17}$, $f_{18}$ \\ 
  $K_{5}$ & $f_{16}$ \\ 
  $K_{6}$ & $f_{19}$ \\   
  \hline
  $L_{2}$ & $f_{9}$ \\ 
  $L_{3}$ & $f_{13}$, $f_{14}$, $f_{15a}$, $f_{15b}$ \\
  \hline 
  \end{tabular}
\end{center}
\caption{This Table shows the operator coefficient(s) contained in each integral of type $J_{n}$, $K_{n}$ or $L_{n}$.}
\label{table1}
\end{table}

In Table \ref{table1} we specify, 
for each integral, which universal coefficients (and correspondingly the operator structures in the universal Lagrangian) it contains. Each operator appears in one specific integral, except $GGG$ (associated to $f_{5}$),
which has a contribution from both $J_{1}$ and $J_{2}$. We note that some operators can be re-written in terms of other operators, and one has to be careful to select a non-redundant basis. This is a point that turns out to be crucial when matching the one-loop effective Lagrangian to a given UV model.

\section{Master Integrals}
\label{app:masterintegrals}
 We introduce for brevity the notation $\text{$\Delta $}m_{a b}^2 \equiv (m_{a}^2-m_{b}^2)$. The complete list~\footnote{Note that one has to perform changes of indices in order to compute all the required integrals.} of the master integrals, $\text{I}[\text{q}^{2\alpha}]\cc{i}{n}\cc{j}{m}\cc{\cdots}{\cdots}\cc{l}{p}$, is given here:
 {\small
\begin{eqnarray*}
\I\cc{i}{2}\cc{j}{1}\cc{k}{1}\cc{l}{1}\cc{m}{\hspace{0.4mm} 1}\cc{n}{1}&=& \frac{i}{4 (4\pi)^2}\Bigg( \Bigg\{ \frac{m_i^2-m_i^2 \ln \left(m_i^2\right) -m_j^2+m_j^2 \ln \left(m_j^2\right)}{\text{$\Delta $}m_{i j}^4 \text{$\Delta $}m_{j k}^2 \text{$\Delta $}m_{j l}^2 \text{$\Delta $}m_{j m}^2 \text{$\Delta $}m_{j n}^2} -(j \leftrightarrow k) +(j \leftrightarrow l) -(j \leftrightarrow m) +(j \leftrightarrow n)  \Bigg\}  \nonumber \\
&& -\frac{\ln \left(m_i^2\right)}{\text{$\Delta $}m_{i j}^2 \text{$\Delta $}m_{i k}^2 \text{$\Delta $}m_{i l}^2 \text{$\Delta $}m_{i m}^2 \text{$\Delta $}m_{i n}^2}  \Bigg) \ , \nonumber \\
\I\cc{i}{2}\cc{j}{1}\cc{k}{1}\cc{l}{1}\cc{m}{1}&=& \frac{i}{3 (4\pi)^2} \Bigg(\Bigg\{\frac{-m_i^2+m_i^2 \ln \left(m_i^2\right)+m_j^2-m_j^2 \ln \left(m_j^2\right)}{\text{$\Delta $}m_{i j}^4 \text{$\Delta $}m_{j k}^2 \text{$\Delta $}m_{j l}^2 \text{$\Delta $}m_{j m}^2} -(j \leftrightarrow k) +(j \leftrightarrow l)  -(j \leftrightarrow m) \Bigg\}\nonumber \\  && + \frac{\ln \left(m_i^2\right)}{\text{$\Delta $}m_{i j}^2 \text{$\Delta $}m_{i k}^2 \text{$\Delta $}m_{i l}^2 \text{$\Delta $}m_{i m}^2} \Bigg) \ , \nonumber \\
\Id\cc{i}{3}\cc{j}{2}\cc{k}{1}\cc{l}{1}&=& -\frac{i}{6 (4\pi)^2 \text{$\Delta $}m_{i j}^8 \text{$\Delta $}m_{i k}^6 \text{$\Delta $}m_{j k}^4 \text{$\Delta $}m_{i l}^6 \text{$\Delta $}m_{j l}^4 \text{$\Delta $}m_{k l}^2}  \Bigg(2 \ln \left(m_l^2\right) m_l^4 \text{$\Delta $}m_{i k}^6 \text{$\Delta $}m_{j k}^4 \text{$\Delta $}m_{i j}^8 \nonumber \\
&& -2 \ln \left(m_i^2\right) \Big(3 m_i^{12}-3 \left(m_k^2+m_l^2\right) m_i^{10}+\left(m_l^4+\left(m_k^2-4 m_j^2\right) m_l^2+m_k^2 \left(m_k^2-4 m_j^2\right)\right) m_i^8 \nonumber \\
&& +m_j^2 \left(2 m_l^4+\left(m_j^2+14 m_k^2\right) m_l^2+m_k^2 \left(m_j^2+2 m_k^2\right)\right) m_i^6-3 m_j^2 m_k^2 m_l^2 \left(m_j^2+2 m_k^2+2 m_l^2\right) m_i^4+ \nonumber \\
&& 2 m_j^2 m_k^4 m_l^4 m_i^2+m_j^4 m_k^4 m_l^4\Big) \text{$\Delta $}m_{j k}^4 \text{$\Delta $}m_{j l}^4 \text{$\Delta $}m_{k l}^2-\text{$\Delta $}m_{i l}^2 \Big(2 \ln \left(m_k^2\right) m_k^4 \text{$\Delta $}m_{i l}^4 \text{$\Delta $}m_{j l}^4 \text{$\Delta $}m_{i j}^8 \nonumber \\
&& +\text{$\Delta $}m_{i k}^2 \big(2 \ln \left(m_j^2\right) m_j^2 \Big(\left(2 m_l^2 m_k^2-m_j^2 m_k^2-m_j^2 m_l^2\right) m_i^2+m_j^2 (\left(3 m_j^2-2 m_k^2\right) m_j^2 \nonumber \\
&& +\left(m_k^2-2 m_j^2\right) m_l^2)\Big) \text{$\Delta $}m_{i k}^4 \text{$\Delta $}m_{i l}^4+\text{$\Delta $}m_{i j}^2 \text{$\Delta $}m_{j k}^2 \big(2 m_j^2 m_i^8+\left(5 m_j^4-9 m_k^2 m_j^2-9 m_l^2 m_j^2+5 m_k^2 m_l^2\right) m_i^6\nonumber \\
&& -\left(\left(3 m_k^2-5 m_j^2\right) m_l^4+\left(2 m_j^4-13 m_k^2 m_j^2+3 m_k^4\right) m_l^2+m_j^2 \left(m_j^4+2 m_k^2 m_j^2-5 m_k^4\right)\right) m_i^4\nonumber \\
&& +\left(m_k^2 \text{$\Delta $}m_{j k}^2 m_j^4+\left(m_j^4-6 m_k^2 m_j^2+m_k^4\right) m_l^4-m_j^2 \left(m_j^4-3 m_k^2 m_j^2+6 m_k^4\right) m_l^2\right) m_i^2\nonumber \\
&& +m_j^2 m_k^2 m_l^2 \left(\left(5 m_k^2-3 m_j^2\right) m_l^2+3 m_j^2 \text{$\Delta $}m_{j k}^2\right)\big) \text{$\Delta $}m_{j l}^2\big) \text{$\Delta $}m_{k l}^2\Big)\Bigg) \, , \nonumber 
\end{eqnarray*}
\begin{eqnarray*}
\Id\cc{i}{2}\cc{j}{2}\cc{k}{2}\cc{l}{1}&=& -\frac{i}{3 (4\pi)^2 \text{$\Delta $}m_{i j}^6 \text{$\Delta $}m_{i k}^6 \text{$\Delta $}m_{i l}^4 \text{$\Delta $}m_{j k}^6 \text{$\Delta $}m_{j l}^4 \text{$\Delta $}m_{k l}^4} \Bigg(m_l^4 \text{$\Delta $}m_{i j}^6 \text{$\Delta $}m_{i k}^6 \text{$\Delta $}m_{j k}^6 \left(-\ln \left(m_l^2\right)\right) \nonumber \\
&& +m_i^2 \ln \left(m_i^2\right) \text{$\Delta $}m_{j k}^6 \text{$\Delta $}m_{j l}^4 \text{$\Delta $}m_{k l}^4 \left(-m_i^4 \left(m_j^2+m_k^2+2 m_l^2\right)-m_i^2 m_j^2 m_k^2+3 m_i^6+2 m_j^2 m_k^2 m_l^2\right) \nonumber \\
&&+\text{$\Delta $}m_{i l}^2 \Big(\text{$\Delta $}m_{i j}^2 \text{$\Delta $}m_{j l}^2 \Big(\text{$\Delta $}m_{i k}^2 \text{$\Delta $}m_{j k}^2 \text{$\Delta $}m_{k l}^2 \Big(m_i^6 \left(m_l^2 \left(m_j^2+m_k^2\right)-2 m_j^2 m_k^2\right)-m_i^4 \Big(m_l^4 \left(m_j^2+m_k^2\right) \nonumber \\
&& +2 m_j^2 m_k^2 m_l^2-2 m_j^2 m_k^2 \left(m_j^2+m_k^2\right)\Big)-m_i^2 \Big(m_l^4 \left(-6 m_j^2 m_k^2+m_j^4+m_k^4\right) \nonumber \\
&& -m_l^2 \left(-2 m_j^4 m_k^2-2 m_j^2 m_k^4+m_j^6+m_k^6\right)+2 m_j^2 m_k^2 \left(-m_j^2 m_k^2+m_j^4+m_k^4\right)\Big)\nonumber \\
&& +m_j^2 m_k^2 m_l^2 \left(-m_l^2 \left(m_j^2+m_k^2\right)+m_j^4+m_k^4\right)\Big)+m_k^2 \text{$\Delta $}m_{i j}^4 \text{$\Delta $}m_{i l}^2 \text{$\Delta $}m_{j l}^2 \ln \left(m_k^2\right) \big(m_i^2 \big(2 m_j^2 m_l^2 \nonumber \\
&& -m_k^2 \left(m_j^2+m_k^2\right)\big)+m_k^4 \left(-m_j^2+3 m_k^2-2 m_l^2\right)\big)\Big)-m_j^2 \text{$\Delta $}m_{i k}^6 \text{$\Delta $}m_{i l}^2 \ln \left(m_j^2\right) \text{$\Delta $}m_{k l}^4 \big(m_j^4 (3 m_j^2-m_k^2 \nonumber \\
&& -2 m_l^2 )-m_i^2 \left(m_j^2 m_k^2+m_j^4-2 m_k^2 m_l^2\right)\big)\Big)\Bigg) \, , \nonumber \\
\I\cc{i}{2}\cc{j}{2}\cc{k}{1}\cc{l}{1}&=& -\frac{i}{3 (4\pi)^2 \text{$\Delta $}m_{i j}^6 \text{$\Delta $}m_{i k}^4 \text{$\Delta $}m_{i l}^4 \text{$\Delta $}m_{j k}^4 \text{$\Delta $}m_{j l}^4 \text{$\Delta $}m_{k l}^2} \Bigg(m_l^2 \text{$\Delta $}m_{i j}^6 \text{$\Delta $}m_{i k}^4 \text{$\Delta $}m_{j k}^4 \left(-\ln \left(m_l^2\right)\right)\nonumber \\
&& +\ln \left(m_i^2\right) \text{$\Delta $}m_{j k}^4 \text{$\Delta $}m_{j l}^4 \text{$\Delta $}m_{k l}^2 \left(-m_i^4 \left(m_j^2+2 \left(m_k^2+m_l^2\right)\right)+m_i^2 m_k^2 m_l^2+3 m_i^6+m_j^2 m_k^2 m_l^2\right)\nonumber \\
&&+\text{$\Delta $}m_{i l}^2 \Big(m_k^2 \text{$\Delta $}m_{i j}^6 \text{$\Delta $}m_{i l}^2 \text{$\Delta $}m_{j l}^4 \ln \left(m_k^2\right)+\text{$\Delta $}m_{i k}^2 \text{$\Delta $}m_{k l}^2 \big(\text{$\Delta $}m_{i j}^2 \text{$\Delta $}m_{j k}^2 \text{$\Delta $}m_{j l}^2 \Big(-m_i^2 \left(m_k^2+m_l^2\right)+m_i^4 \nonumber \\
&&-m_j^2 \left(m_k^2+m_l^2\right)+m_j^4+2 m_k^2 m_l^2\Big)+\text{$\Delta $}m_{i k}^2 \text{$\Delta $}m_{i l}^2 \ln \left(m_j^2\right) (-m_j^4 \left(m_i^2+2 \left(m_k^2+m_l^2\right)\right) \nonumber \\
&& +m_i^2 m_k^2 m_l^2+m_j^2 m_k^2 m_l^2+3 m_j^6)\big)\Big)\Bigg) \, , \nonumber \\
\I\cc{i}{2}\cc{j}{1}\cc{k}{1}\cc{l}{1}&=& \frac{i}{(4\pi)^2} \Bigg( \Bigg\{\frac{\text{$\Delta $}m_{i j}^2-m_i^2 \ln \left(m_i^2\right)+m_j^2 \ln \left(m_j^2\right)}{\text{$\Delta $}m_{i j}^4 \text{$\Delta $}m_{j k}^2 \text{$\Delta $}m_{j l}^2}  -(j \leftrightarrow k) +(j \leftrightarrow l) \Bigg\} \nonumber \\
&&+ \frac{\ln \left(m_i^2\right)}{\text{$\Delta $}m_{i j}^2 \text{$\Delta $}m_{i k}^2 \text{$\Delta $}m_{i l}^2} \Bigg) \, , \nonumber \\
\Id\cc{i}{4}\cc{j}{1}\cc{k}{1}&=& \frac{i}{12 (4\pi)^2 m_i^2 \text{$\Delta $}m_{i j}^8 \text{$\Delta $}m_{i k}^8 \text{$\Delta $}m_{j k}^2} \Bigg(\text{$\Delta $}m_{i j}^2 \text{$\Delta $}m_{i k}^2 \text{$\Delta $}m_{j k}^2 \Big(5 m_i^6 \left(m_j^2+m_k^2\right)-m_i^4 \left(22 m_j^2 m_k^2+m_j^4+m_k^4\right) \nonumber \\
&& +5 m_i^2 m_j^2 m_k^2 \left(m_j^2+m_k^2\right)+2 m_i^8+2 m_j^4 m_k^4\Big)-6 m_i^2 m_k^4 \text{$\Delta $}m_{i j}^8 \ln \left(m_k^2\right)+6 m_i^2 m_j^4 \text{$\Delta $}m_{i k}^8 \ln \left(m_j^2\right) \nonumber \\
&& -6 m_i^2 \ln \left(m_i^2\right) \text{$\Delta $}m_{j k}^2 \left(m_i^4-m_j^2 m_k^2\right) \left(m_i^4 \left(m_j^2+m_k^2\right)-4 m_i^2 m_j^2 m_k^2+m_j^2 m_k^2 \left(m_j^2+m_k^2\right)\right)\Bigg) \, , \nonumber \\
\Id\cc{i}{3}\cc{j}{2}\cc{k}{1}&=& \frac{i}{4 (4\pi)^2 \text{$\Delta $}m_{i j}^8 \text{$\Delta $}m_{i k}^6 \text{$\Delta $}m_{j k}^4}  \Bigg(-\text{$\Delta $}m_{i j}^2 \text{$\Delta $}m_{i k}^2 \text{$\Delta $}m_{j k}^2 \Big(m_k^4 \left(m_i^2+5 m_j^2\right)-3 m_k^2 \left(m_i^2+m_j^2\right){}^2 \nonumber \\
&& +m_i^2 m_j^2 \left(5 m_i^2+m_j^2\right)\Big)-2 m_k^4 \text{$\Delta $}m_{i j}^8 \ln \left(m_k^2\right)-2 m_j^2 \text{$\Delta $}m_{i k}^6 \ln \left(m_j^2\right) \Big(m_j^2 \left(m_i^2+2 m_j^2\right) \nonumber \\
&& -m_k^2 \left(2 m_i^2+m_j^2\right)\Big)+2 \ln \left(m_i^2\right) \text{$\Delta $}m_{j k}^4 \left(-6 m_i^4 m_j^2 m_k^2+m_j^2 m_k^4 \left(2 m_i^2+m_j^2\right)+m_i^6 \left(m_i^2+2 m_j^2\right)\right)\Bigg) \, ,\nonumber 
\end{eqnarray*}  
\begin{eqnarray*}
\I\cc{i}{3}\cc{j}{1}\cc{k}{1}&=& \frac{i}{4 (4\pi)^2 m_i^2 \text{$\Delta $}m_{i j}^6 \text{$\Delta $}m_{i k}^6 \text{$\Delta $}m_{j k}^2} \Bigg(\text{$\Delta $}m_{i j}^2 \text{$\Delta $}m_{i k}^2 \text{$\Delta $}m_{j k}^2 \left(m_i^2 \left(m_j^2+m_k^2\right)-3 m_i^4+m_j^2 m_k^2\right) \nonumber \\
&& +2 m_i^2 \Big(m_k^2 \text{$\Delta $}m_{i j}^6 \ln \left(m_k^2\right)-m_j^2 \text{$\Delta $}m_{i k}^6 \ln \left(m_j^2\right)+\ln \left(m_i^2\right) \text{$\Delta $}m_{j k}^2 \big(-3 m_i^2 m_j^2 m_k^2+m_i^6 \nonumber \\
&&+m_j^2 m_k^2 \left(m_j^2+m_k^2\right) \big)\Big)\Bigg) \, , \nonumber \\
\Id\cc{i}{2}\cc{j}{2}\cc{k}{2}&=& \frac{i}{2  (4\pi)^2 \text{$\Delta $}m_{i j}^6 \text{$\Delta $}m_{i k}^6 \text{$\Delta $}m_{j k}^6} \Bigg(\text{$\Delta $}m_{i j}^2 \text{$\Delta $}m_{i k}^2 \text{$\Delta $}m_{j k}^2 \Big(m_k^4 \left(m_i^2+m_j^2\right)+m_k^2 \left(-6 m_i^2 m_j^2+m_i^4+m_j^4\right) \nonumber \\
&& +m_i^2 m_j^2 \left(m_i^2+m_j^2\right)\Big)+2 m_k^2 \text{$\Delta $}m_{i j}^6 \ln \left(m_k^2\right) \left(m_k^4-m_i^2 m_j^2\right)+2 m_j^2 \text{$\Delta $}m_{i k}^6 \ln \left(m_j^2\right) \left(m_j^4-m_i^2 m_k^2\right) \nonumber \\
&&-2 m_i^2 \ln \left(m_i^2\right) \text{$\Delta $}m_{j k}^6 \left(m_i^4-m_j^2 m_k^2\right)\Bigg) \, , \nonumber  \\
 \I\cc{i}{2}\cc{j}{2}\cc{k}{1}&=&  \frac{i}{2 (4\pi)^2 \text{$\Delta $}m_{i j}^6 \text{$\Delta $}m_{i k}^4 \text{$\Delta $}m_{j k}^4}  \Bigg(\text{$\Delta $}m_{i j}^2 \text{$\Delta $}m_{i k}^2 \text{$\Delta $}m_{j k}^2 \left(m_i^2+m_j^2-2 m_k^2\right)+m_k^2 \text{$\Delta $}m_{i j}^6 \ln \left(m_k^2\right) \nonumber \\
 && -\text{$\Delta $}m_{i k}^4 \ln \left(m_j^2\right) \left(m_k^2 \left(m_i^2+m_j^2\right)-2 m_j^4\right)+\ln \left(m_i^2\right) \text{$\Delta $}m_{j k}^4 \left(m_k^2 \left(m_i^2+m_j^2\right)-2 m_i^4\right)\Bigg) \, , \nonumber \\ 
  \I\cc{i}{2}\cc{j}{1}\cc{k}{1}&=&  \frac{i \left(m_k^2 \text{$\Delta $}m_{i j}^4 \left(-\ln \left(m_k^2\right)\right)+m_j^2 \text{$\Delta $}m_{i k}^4 \ln \left(m_j^2\right)+\text{$\Delta $}m_{j k}^2 \left(\text{$\Delta $}m_{i j}^2 \text{$\Delta $}m_{i k}^2+\ln \left(m_i^2\right) \left(m_j^2 m_k^2-m_i^4\right)\right)\right)}{(4\pi)^2 \text{$\Delta $}m_{i j}^4 \text{$\Delta $}m_{i k}^4 \text{$\Delta $}m_{j k}^2} \, , \nonumber \\
\Iq\cc{i}{5}\cc{j}{2}&=& \frac{i \left(-12 m_i^6 m_j^2-36 m_i^4 m_j^4+44 m_i^2 m_j^6+m_i^8+3 m_j^8+12 m_i^2 m_j^4 \left(3 m_i^2+2 m_j^2\right) \ln \left(\frac{m_i^2}{m_j^2}\right)\right)}{24 (4\pi)^2 m_i^2 \text{$\Delta $}m_{i j}^{12}} \, , \nonumber \\
\Id\cc{i}{4}\cc{j}{2}&=& \frac{i \left(9 m_i^4 m_j^2-9 m_i^2 m_j^4+m_i^6-m_j^6 -6 m_i^2 m_j^2 \left(m_i^2+m_j^2\right) \ln \left(\frac{m_i^2}{m_j^2}\right)\right)}{6 (4\pi)^2 m_i^2 \text{$\Delta $}m_{i j}^{10}} \, ,\nonumber \\
 \Id\cc{i}{4}\cc{j}{1}&=& -\frac{i \left(-6 m_i^4 m_j^2+3 m_i^2 m_j^4+m_i^6+2 m_j^6 +6 m_i^2 m_j^4 \ln \left(\frac{m_i^2}{m_j^2}\right)\right)}{6 (4\pi)^2 m_i^2 \text{$\Delta $}m_{i j}^8} \, , \nonumber \\
\Id\cc{i}{3}\cc{j}{3}&=& \frac{i \left(-3 m_i^4+3 m_j^4 + \left(4 m_i^2 m_j^2+m_i^4+m_j^4\right) \ln \left(\frac{m_i^2}{m_j^2}\right)\right)}{2 (4\pi)^2 \text{$\Delta $}m_{i j}^{10}}  \, ,\nonumber \\
 \Id \cc{i}{3}\cc{j}{2}&=& -\frac{i \left(m_i^4-5 m_j^4 + 4 m_i^2 m_j^2+2 m_j^2 \left(2 m_i^2+m_j^2\right) \ln \left(\frac{m_j^2}{m_i^2}\right) \right)}{2 (4\pi)^2 \text{$\Delta $}m_{i j}^8} \, , \nonumber \\
\I\cc{i}{3}\cc{j}{2}&=&  \frac{i \left(4 m_i^2 m_j^2-5 m_i^4+m_j^4-2 m_i^2 \left(m_i^2+2 m_j^2\right) \ln \left(\frac{m_j^2}{m_i^2}\right)\right)}{4 (4\pi)^2 m_i^2 \text{$\Delta $}m_{i j}^8} \, , \nonumber \\
\I\cc{i}{3}\cc{j}{1}&=& -\frac{i \left(m_i^4-m_j^4 + 2 m_i^2 m_j^2 \ln \left(\frac{m_j^2}{m_i^2}\right) \right)}{2 (4\pi)^2 m_i^2 \text{$\Delta $}m_{i j}^6}  \, ,\nonumber 
\quad \quad \quad \quad
\I\cc{i}{2}\cc{j}{2} =  -\frac{i \left( -2 m_i^2+2 m_j^2 + \left(m_i^2+m_j^2\right) \ln \left(\frac{m_i^2}{m_j^2}\right)\right)}{(4\pi)^2 \text{$\Delta $}m_{i j}^6}  \, ,\nonumber \\
\Iq\cc{i}{6}&=& -\frac{i}{20 (4\pi)^2 m_i^4}   \, ,\nonumber 
\quad \quad \quad \quad \quad \quad \quad \quad \quad \quad \quad \quad \quad
\Id\cc{i}{5} = -\frac{i}{12 (4\pi)^2 m_i^4}  \, ,\nonumber 
\quad  \quad 
\I\cc{i}{4} = -\frac{i}{6 (4\pi)^2 m_i^4} \, .
\label{eq:fexpressions2}
\end{eqnarray*}

} 

\newpage
\section{Mass Dependences of the Universal Coefficients}
\label{app:fexplicit}

In this Appendix, we give the explicit expressions for the $f_{N}^{ijk...}$ coefficients as function of several masses. To reduce the length as much as possible, we use the following notation: $ \tilde{f}_{N}^{ijk...} \equiv -i(4\pi)^2  f_{N}^{ijk...} $ and $\text{$\Delta $}m_{a b}^2 \equiv (m_{a}^2-m_{b}^2)$. The expressions for $f_{15a}^{ijk}$, $f_{15b}^{ijk}$, $f_{17}^{ijkl}$ and $f_{18}^{ijkl}$ are quite lengthy so we do not write them explicitly here. However, they are straightforward to obtain by plugging the master integrals of Appendix B into (\ref{eq:fexpressions}). 
The mass dependences of the universal coefficients, $\tilde{f}_{N}^{ijk...}$,  are as follows: 

{\small 

\begin{eqnarray*}
\tilde{f}_{5}^{ij} &=&  \frac{6 m_i^4 m_j^2-3 m_i^2 m_j^4-6 m_i^2 m_j^4 \ln \left(\frac{m_i^2}{m_j^2}\right)-m_i^6-2 m_j^6}{54 \text{$\Delta $}m_{i j}^8}+\frac{1}{270} \left(\frac{7}{m_i^2}-\frac{9}{m_j^2}\right)  \, ,\nonumber \\
 \tilde{f}_{6}^{ij} &=&  \frac{7}{180 m_i^2}-\frac{1}{20 m_j^2} \, , \nonumber \\
\tilde{f}_{7}^{ij} &=& \frac{-2 m_i^4 m_j^2+\frac{5}{2} m_i^2 m_j^4+2 m_i^4 m_j^2 \ln \left(\frac{m_i^2}{m_j^2}\right)+m_i^2 m_j^4 \ln \left(\frac{m_i^2}{m_j^2}\right)-\frac{m_i^6}{2}}{\text{$\Delta $}m_{i j}^8}  \, ,\nonumber \\
\tilde{f}_{8}^{ijk} &=& \frac{m_i^2 m_j^2 m_k^2+m_i^2 m_j^2 m_k^2 \ln \left(m_i^2\right)-m_i^4 m_j^2-m_i^4 m_k^2+m_i^6-m_i^6 \ln \left(m_i^2\right)}{\text{$\Delta $}m_{i j}^4 \text{$\Delta $}m_{i k}^4} \nonumber \\
&& + \frac{m_i^2 m_j^2 \ln \left(m_j^2\right)}{\text{$\Delta $}m_{i j}^4 \text{$\Delta $}m_{j k}^2} -\frac{m_i^2 m_k^2 \ln \left(m_k^2\right)}{\text{$\Delta $}m_{i k}^4 \text{$\Delta $}m_{j k}^2}  \, ,\nonumber \\
 \tilde{f}_{9}^{ij} &=& \frac{m_i^4 m_j^2+m_i^2 m_j^4+2 m_i^4 m_j^2 \ln \left(\frac{m_i^2}{m_j^2}\right)-2 m_i^2 m_j^4 \ln \left(\frac{m_i^2}{m_j^2}\right)-m_i^6-m_j^6}{4 \text{$\Delta $}m_{i j}^8}  \, ,\nonumber \\
\tilde{f}_{10}^{ijkl} &=& \left\{ \frac{-m_i^2 m_j^2+m_i^2 m_j^2 \ln \left(m_j^2\right)+m_i^4-m_i^4 \ln \left(m_i^2\right)}{2 \text{$\Delta $}m_{i j}^4 \text{$\Delta $}m_{j k}^2 \text{$\Delta $}m_{j l}^2} + \left( j \leftrightarrow k \right) - \left( j \leftrightarrow l \right) \right\} \nonumber \\
&&+ \frac{m_i^2 \ln \left(m_i^2\right)}{2 \text{$\Delta $}m_{i j}^2 \text{$\Delta $}m_{i k}^2 \text{$\Delta $}m_{i l}^2}  \, ,\nonumber \\
\tilde{f}_{11}^{ijk} &=& \frac{1}{2 \text{$\Delta $}m_{i j}^2 \text{$\Delta $}m_{i k}^6 \text{$\Delta $}m_{j k}^6} \Bigg( 3 m_i^2 m_j^4 m_k^4-3 m_i^4 m_j^2 m_k^4+6 m_i^2 m_j^2 m_k^6 \ln \left(\frac{m_i^2}{m_j^2}\right)-6 m_i^2 m_j^4 m_k^4 \ln \left(\frac{m_i^2}{m_k^2}\right) \nonumber \\
&& +6 m_i^4 m_j^2 m_k^4 \ln \left(\frac{m_j^2}{m_k^2}\right)+2 m_i^2 m_j^6 m_k^2 \ln \left(\frac{m_i^2}{m_k^2}\right)-2 m_i^6 m_j^2 m_k^2 \ln \left(\frac{m_j^2}{m_k^2}\right)-m_i^4 m_j^6+m_i^6 m_j^4 \nonumber \\
&& -3 m_i^2 m_k^8+4 m_i^4 m_k^6-m_i^6 m_k^4-2 m_i^2 m_k^8 \ln \left(\frac{m_i^2}{m_k^2}\right)+3 m_j^2 m_k^8-4 m_j^4 m_k^6+m_j^6 m_k^4+2 m_j^2 m_k^8 \ln \left(\frac{m_j^2}{m_k^2}\right) \Bigg)  \, ,\nonumber \\
 \tilde{f}_{12,a}^{ij} &=& \frac{28 m_i^6 m_j^2-28 m_i^2 m_j^6-12 m_i^6 m_j^2 \ln \left(\frac{m_i^2}{m_j^2}\right)-36 m_i^4 m_j^4 \ln \left(\frac{m_i^2}{m_j^2}\right)-12 m_i^2 m_j^6 \ln \left(\frac{m_i^2}{m_j^2}\right)+m_i^8-m_j^8}{36 \text{$\Delta $}m_{i j}^{12}}  \, ,\nonumber \\
  \tilde{f}_{12,b}^{ij} &=&   \tilde{f}_{12,a}^{ij} \, , \nonumber
\end{eqnarray*}
\begin{eqnarray*}
\tilde{f}_{12,c}^{ij} &=& \frac{26 m_i^6 m_j^2-27 m_i^4 m_j^4-2 m_i^2 m_j^6-15 m_i^6 m_j^2 \ln \left(\frac{m_i^2}{m_j^2}\right)-18 m_i^4 m_j^4 \ln \left(\frac{m_i^2}{m_j^2}\right)+3 m_i^2 m_j^6 \ln \left(\frac{m_i^2}{m_j^2}\right)+2 m_i^8+m_j^8}{18 \text{$\Delta $}m_{i j}^{12}}  \, ,\nonumber \\
 \tilde{f}_{13}^{ijk} &=& \frac{1}{8 \text{$\Delta $}m_{i j}^6 \text{$\Delta $}m_{i k}^6 \text{$\Delta $}m_{j k}^6}  \Bigg(m_j^4 m_i^{10}-m_k^4 m_i^{10}\nonumber \\
 && -2 \ln \left(\frac{m_j^2}{m_k^2}\right) m_j^2 m_k^2 m_i^{10}+2 \ln \left(\frac{m_i^2}{m_j^2}\right) m_j^6 m_i^8-6 m_j^6 m_i^8-2 \ln \left(\frac{m_i^2}{m_k^2}\right) m_k^6 m_i^8+7 m_k^6 m_i^8 \\
 && +6 \ln \left(m_i^2\right) m_j^2 m_k^4 m_i^8+4 \ln \left(m_j^2\right) m_j^2 m_k^4 m_i^8-10 \ln \left(m_k^2\right) m_j^2 m_k^4 m_i^8-10 m_j^2 m_k^4 m_i^8 \\
 && -6 \ln \left(m_i^2\right) m_j^4 m_k^2 m_i^8+8 \ln \left(m_j^2\right) m_j^4 m_k^2 m_i^8-2 \ln \left(m_k^2\right) m_j^4 m_k^2 m_i^8+9 m_j^4 m_k^2 m_i^8+7 m_j^8 m_i^6 \\
 && -2 \ln \left(\frac{m_i^2}{m_k^2}\right) m_k^8 m_i^6-7 m_k^8 m_i^6+6 \ln \left(\frac{m_i^2}{m_k^2}\right) m_j^2 m_k^6 m_i^6-6 \ln \left(m_i^2\right) m_j^4 m_k^4 m_i^6-24 \ln \left(m_j^2\right) m_j^4 m_k^4 m_i^6 \\
 && +30 \ln \left(m_k^2\right) m_j^4 m_k^4 m_i^6+8 m_j^4 m_k^4 m_i^6+2 \ln \left(m_i^2\right) m_j^6 m_k^2 m_i^6+4 \ln \left(m_j^2\right) m_j^6 m_k^2 m_i^6-6 \ln \left(m_k^2\right) m_j^6 m_k^2 m_i^6 \\
 && -8 m_j^6 m_k^2 m_i^6-2 m_j^{10} m_i^4+m_k^{10} m_i^4+10 \ln \left(m_i^2\right) m_j^2 m_k^8 m_i^4-4 \ln \left(m_j^2\right) m_j^2 m_k^8 m_i^4-6 \ln \left(m_k^2\right) m_j^2 m_k^8 m_i^4 \\
 && +10 m_j^2 m_k^8 m_i^4-30 \ln \left(m_i^2\right) m_j^4 m_k^6 m_i^4+24 \ln \left(m_j^2\right) m_j^4 m_k^6 m_i^4+6 \ln \left(m_k^2\right) m_j^4 m_k^6 m_i^4-8 m_j^4 m_k^6 m_i^4 \\
 && +30 \ln \left(\frac{m_i^2}{m_k^2}\right) m_j^6 m_k^4 m_i^4-10 \ln \left(\frac{m_i^2}{m_k^2}\right) m_j^8 m_k^2 m_i^4-m_j^8 m_k^2 m_i^4-2 \ln \left(\frac{m_i^2}{m_j^2}\right) m_j^2 m_k^{10} m_i^2 \\
 && +2 \ln \left(m_i^2\right) m_j^4 m_k^8 m_i^2-8 \ln \left(m_j^2\right) m_j^4 m_k^8 m_i^2+6 \ln \left(m_k^2\right) m_j^4 m_k^8 m_i^2-9 m_j^4 m_k^8 m_i^2+6 \ln \left(m_i^2\right) m_j^6 m_k^6 m_i^2 \\
 && -4 \ln \left(m_j^2\right) m_j^6 m_k^6 m_i^2-2 \ln \left(m_k^2\right) m_j^6 m_k^6 m_i^2+8 m_j^6 m_k^6 m_i^2-10 \ln \left(\frac{m_i^2}{m_k^2}\right) m_j^8 m_k^4 m_i^2+m_j^8 m_k^4 m_i^2 \\
 && +4 \ln \left(\frac{m_i^2}{m_k^2}\right) m_j^{10} m_k^2 m_i^2-m_j^4 m_k^{10}+2 \ln \left(\frac{m_j^2}{m_k^2}\right) m_j^6 m_k^8+6 m_j^6 m_k^8-7 m_j^8 m_k^6+2 m_j^{10} m_k^4 \Bigg)  \, ,
\end{eqnarray*}
 \begin{eqnarray*}
\tilde{f}_{14}^{ijk} &=& \frac{1}{12 \text{$\Delta $}m_{i j}^8 \text{$\Delta $}m_{i k}^2 \text{$\Delta $}m_{j k}^8} \Bigg(-24 m_i^2 m_j^8 m_k^4+24 m_i^4 m_j^8 m_k^2+32 m_i^2 m_j^6 m_k^6-32 m_i^6 m_j^6 m_k^2-6 m_i^2 m_j^4 m_k^8 \\
&& -24 m_i^4 m_j^4 m_k^6+24 m_i^6 m_j^4 m_k^4+6 m_i^8 m_j^4 m_k^2+3 m_i^4 m_j^2 m_k^8-3 m_i^8 m_j^2 m_k^4+24 m_i^2 m_j^8 m_k^4 \ln \left(\frac{m_j^2}{m_k^2}\right) \\
&& +24 m_i^4 m_j^8 m_k^2 \ln \left(\frac{m_i^2}{m_j^2}\right)-36 m_i^4 m_j^6 m_k^4 \ln \left(\frac{m_i^2}{m_k^2}\right)+24 m_i^4 m_j^4 m_k^6 \ln \left(\frac{m_i^2}{m_j^2}\right)+24 m_i^6 m_j^4 m_k^4 \ln \left(\frac{m_j^2}{m_k^2}\right)\\
&& -6 m_i^4 m_j^2 m_k^8 \ln \left(\frac{m_i^2}{m_j^2}\right)  -6 m_i^8 m_j^2 m_k^4 \ln \left(\frac{m_j^2}{m_k^2}\right)-2 m_i^2 m_j^{12}-3 m_i^4 m_j^{10}+6 m_i^6 m_j^8-m_i^8 m_j^6\\
&& -6 m_i^4 m_j^{10} \ln \left(\frac{m_i^2}{m_j^2}\right)+2 m_i^6 m_k^8-2 m_i^8 m_k^6+2 m_j^{12} m_k^2+3 m_j^{10} m_k^4-6 m_j^8 m_k^6+m_j^6 m_k^8-6 m_j^{10} m_k^4 \ln \left(\frac{m_j^2}{m_k^2}\right) \Bigg)  \, ,\nonumber \\
\tilde{f}_{15a}^{ijk} &=&  \text{see (\ref{eq:fexpressions}) and Appendix B } \, , \nonumber \\
\tilde{f}_{15b}^{ijk} &=&  \text{see (\ref{eq:fexpressions}) and Appendix B} \, , \nonumber \\ 
\end{eqnarray*}
 \begin{eqnarray}
 \tilde{f}_{16}^{ijklm} &=& \Bigg\{ \frac{m_i^2 m_j^2-m_i^2 m_j^2 \ln \left(m_j^2\right)-m_i^4+m_i^4 \ln \left(m_i^2\right)}{3 \text{$\Delta $}m_{i j}^4 \text{$\Delta $}m_{j k}^2 \text{$\Delta $}m_{j l}^2 \text{$\Delta $}m_{j m}^2} - \left( j \leftrightarrow k \right) + \left( j \leftrightarrow l \right) + \left( j \leftrightarrow m \right) \Bigg\} \nonumber \\
 &&  -\frac{m_i^2 \ln \left(m_i^2\right)}{3 \text{$\Delta $}m_{i j}^2 \text{$\Delta $}m_{i k}^2 \text{$\Delta $}m_{i l}^2 \text{$\Delta $}m_{i m}^2}  \, ,\nonumber \\
\tilde{f}_{17}^{ijkl} &=&  \text{see (\ref{eq:fexpressions}) and Appendix B}  \, ,\nonumber \\ 
\tilde{f}_{18}^{ijkl} &=&  \text{see (\ref{eq:fexpressions}) and Appendix B}  \, ,\nonumber \\ 
\tilde{f}_{19}^{ijklmn} &=& \Bigg\{ \frac{m_i^2 m_j^2-m_i^2 m_j^2 \ln \left(m_j^2\right)-m_i^4+m_i^4 \ln \left(m_i^2\right)}{4 \text{$\Delta $}m_{i j}^4 \text{$\Delta $}m_{j k}^2 \text{$\Delta $}m_{j l}^2 \text{$\Delta $}m_{j m}^2 \text{$\Delta $}m_{j n}^2} + \left( j \leftrightarrow k \right) - \left( j \leftrightarrow l \right) + \left( j \leftrightarrow m \right) - \left( j \leftrightarrow n \right) \Bigg\} \nonumber \\
 &&  -\frac{m_i^2 \ln \left(m_i^2\right)}{4 \text{$\Delta $}m_{i j}^2 \text{$\Delta $}m_{i k}^2 \text{$\Delta $}m_{i l}^2 \text{$\Delta $}m_{i m}^2 \text{$\Delta $}m_{i n}^2}  \, . \nonumber \\
\end{eqnarray}

} 

We give now the expressions of the full coefficients $f_{N}^{ijk...}$ coefficients as appearing in (\ref{eq:universallagrangian}) in the case where all masses are equals, $m_{i}=m_{j}= \cdots =m$. In this degenerate limit, these coefficients read:
\begin{eqnarray}
 f_{5} =  -\frac{i}{(4\pi)^{2} 60 m^2} , \quad & f_{11} =  \frac{i}{(4\pi)^{2} 12 m^4} , \quad & f_{15a} = \frac{i}{(4\pi)^{2} 60 m^4}  \, ,\nonumber \\
 f_{6} =  -\frac{i}{(4\pi)^{2}90 m^2} , \quad & f_{12,a} =  0 , \quad & f_{15b}= \frac{i}{(4\pi)^{2} 60 m^4}  \, ,\nonumber \\ 
 f_{7} =  -\frac{i}{(4\pi)^{2}12 m^2} , \quad & f_{12,b} =  0 , \quad & f_{16} = -\frac{i}{(4\pi)^2 60 m^6} \nonumber \\ 
 f_{8} =  -\frac{i}{(4\pi)^{2}6 m^2} , \quad & f_{12,c} =  \frac{i}{(4\pi)^{2} 120 m^4} , \quad & f_{17} = -\frac{i}{(4\pi)^2 20 m^6}  \, , \nonumber \\ 
 f_{9} =  -\frac{i}{(4\pi)^{2}12 m^2} , \quad & f_{13} =  \frac{i}{(4\pi)^{2} 24 m^4}  , \quad & f_{18} = -\frac{i}{(4\pi)^2 30 m^6} \, , \nonumber \\ 
 f_{10} =  \frac{i}{(4\pi)^{2}24 m^4} , \quad & f_{14} = \frac{-i}{(4\pi)^{2} 60 m^4} , \quad & f_{19} =  \frac{i}{(4\pi)^2 120 m^8}   \, .
\end{eqnarray}
 %

\section{Application: Integrating out Squarks}
\label{app:stop}

The relative simplicity of the approach presented in this work is illustrated by integrating out squarks in the MSSM,
whose leading-order contribution necessarily appears at one-loop order due to R-parity. Currently, gluon fusion 
(which occurs at one-loop order in the SM) provides the strongest constraint on any dimension-6 operator in the Higgs sector. 
The dimension-6 operators contributing to the $h\gamma\gamma$ coupling are also well constrained. This is why in~\cite{DEQY1} our phenomenological studies focused on mainly these two (combinations of) dimension-6 operators as well as the well known $S$ and $T$~\cite{PeskinTakeuchi} (or equivalently $\epsilon_{1,2}$~\cite{AltarelliBarbieri}) parameters from electroweak precision measurements. In this Appendix, in order to demonstrate and validate the general method presented in this work, we compute the full set of  Wilson coefficients for the bosonic operators involving the Higgs in the SM EFT, listed in Table~\ref{dim6-operators}, which supplement the SM Lagrangian as depicted in the following Lagrangian,
\begin{equation}
{\cal L}_{\text{SM-EFT}} = {\cal L}_{\text{SM}} + \sum_i c_i{\cal O}_i \, ,
\end{equation}
${\cal O}_{i}$ being the dimension-6 operators, $c_{i}$ their associated Wilson coefficients and with the multiplet masses the cutoff scales of the EFT.
\renewcommand\arraystretch{1.4}
\begin{table}[!h]
\centering
\begin{tabular}{|rcl|rcl|}\hline
 \({\cal O}_{GG}\) &\(=\)& \(g_s^2 \abs{H}^2G_{\mu \nu }^aG^{a,\mu \nu }\) & \({\cal O}_H\)   &\(=\)& \(\frac{1}{2}\big(\partial_{\mu} \abs{H}^2\big)^2\)\\
 \({\cal O}_{WW}\) &\(=\)& \(g^2  \abs{H}^2 W_{\mu \nu }^aW^{a,\mu \nu } \) &  \({\cal O}_T\)   &\(=\)& \(\frac{1}{2}\big( H^{\dag} \Dfbd H\big)^2\) \\
 \({\cal O}_{BB}\) &\(=\)& \(g'^2 \abs{H}^2 B_{\mu \nu }B^{\mu \nu }\) & \({\cal O}_R\)   &\(=\)& \(\abs{H}^2\abs{D_{\mu}H}^2\) \\
 \({\cal O}_{WB}\) &\(=\)& \(2gg'H^\dag {t^a}H W_{\mu \nu }^a B^{\mu \nu }\) &  \({\cal O}_D\)   &\(=\)& \(\abs{D^2H}^2\) \\
 \({\cal O}_W\)   &\(=\)& \(ig\big(H^\dag t^a \Dfbu H\big)D^\nu W_{\mu \nu }^a\) &  \({\cal O}_6\)   &\(=\)& \(\abs{H}^6\) \\
 \({\cal O}_B\)   &\(=\)& \(ig'Y_H\big(H^\dag \Dfbu H\big)\partial^\nu B_{\mu \nu }\)  &   & & \\
%
%
  \hline
\end{tabular}
\caption{\label{dim6-operators} Dimension-6 SM EFT CP-even bosonic operators involving the Higgs field.}
\vspace{-10pt}
\end{table}
\renewcommand\arraystretch{0}

The $M$ and $U$ matrices are given by the quadratic stop term in the MSSM Lagrangian,
\begin{equation*}
\mathcal{L}_\text{MSSM} \supset \Phi^\dagger (M^2 + U(x)) \Phi	\, ,
\end{equation*}
where $ \Phi = (\tilde{Q} \,, \tilde{t}_R^*)$, and 
\begin{equation*}
M^2 = \twomatrix{ m^2_{\tilde{Q}} & 0 \\ 0 & m^2_{\tilde{t}_R} } \, , 
\end{equation*}
\begin{equation}
U =\twomatrix{
(h_t^2 + \frac{1}{2}g_2^2c^2_\beta)\tilde{H}\tilde{H}^\dagger+\frac{1}{2}g_2^2 s^2_\beta HH^\dagger - \frac{1}{2}(g_1^2 Y_{\tilde{Q}} c_{2\beta} + \frac{1}{2} g_2^2)|H|^2 
&
h_t X_t \tilde{H} 
\\ 
h_t X_t \tilde{H}^\dagger
&
(h_t^2 - \frac{1}{2}g_1^2 Y_{\tilde{t}_R} c_{2\beta})|H|^2
} 	\, .
\label{eq:MSSMU}
\end{equation}
Here we have defined $\tilde{H} \equiv i\sigma^2 H^*$, $h_t \equiv y_t s_\beta$, $X_t \equiv A_t - \mu \cot\beta$, 
and the hypercharges are $Y_{\tilde{Q}} = 1/6, Y_{\tilde{t}_R} = -2/3$. The MSSM mass matrix entries $m_{\tilde{Q}}$ and $m_{\tilde{t}_R}$ denote the soft supersymmetry-breaking masses. We note that 
$\tilde{Q}=\left(\tilde{t}_{L},\tilde{b}_{L}\right)$ is an $SU(2)_L$ doublet, so $U$ is implicitly a $3 \times 3$ matrix with an additional trace over color quantum number. In the case of $SU(3)_c$, $G_{\mu\nu}$ is the usual gluon field strength. In the case of $SU(2)_L$, the field strength reads
\begin{equation}
G^\prime_{\mu\nu} = \twomatrix{
{W^\prime}^a_{\mu\nu}\tau^a + Y_{\tilde{Q}}B^\prime_{\mu\nu}\identity_{2 \times 2} & 0 \\
0 & -Y_{\tilde{t}_R}B^\prime_{\mu\nu}
}	\, .
\label{eq:MSSMG}
\end{equation}
Plugging all of this into (\ref{eq:universallagrangian}), summing over the matrix indices, and rearranging the resulting terms into the dimension-6 operators of Table~\ref{dim6-operators}, we obtain the Wilson coefficients that we list below at the end of the appendix. For convenience the expressions have been multiplied by a factor $(4\pi)^2$ and we defined ${\bar X_t} \equiv h_t X_t$. 

It is instructive to see how each Wilson coefficient originates from the combinations of terms in the universal one-loop effective Lagrangian. We list in Table~\ref{tab:fcontribc} the contributions from each universal coefficient to the Wilson coefficient of the dimension-6 operators listed in Table~\ref{dim6-operators}. These are further separated into the specific term with a certain power of $X_t$ that the universal coefficient is responsible for. Looking at each entry in the matrix $U$ and $G^\prime_{\mu\nu}$ of Eqs.~\ref{eq:MSSMU} and \ref{eq:MSSMG} respectively, we can see immediately for some terms how this leads to the operators in Table.~\ref{dim6-operators} when plugged in to each term of Eq.~\ref{eq:universallagrangian}, each with its own universal coefficient. Other terms in the universal one-loop effective Lagrangian may involve more manipulations than others to obtain the final set of operators in a non-redundant basis.

\begin{table}[h!]
\begin{center}
\begin{tabular}{|M{2cm}|M{2cm}|M{2cm}|M{2cm}|M{2cm}|N}
  \hline
 & $X_{t}^{0}$ & $X_{t}^{2}$ & $X_{t}^{4}$ & $X_{t}^{6}$ &\\[20pt]
  \hline  
  $c_{6}$ & $f_{8}$ & $f_{10}$ & $f_{16}$ & $f_{19}$ &\\[20pt]
    \hline 
  $c_{H}$ & $f_{7}$ & $f_{11}$ & $f_{17}$, $f_{18}$ & - &\\[20pt]
    \hline 
  $c_{T}$ & $f_{7}$ & $f_{11}$ & $f_{17}$, $f_{18}$ & - &\\[20pt]
    \hline 
  $c_{R}$ & $f_{7}$ & $f_{11}$ & $f_{17}$ & - &\\[20pt]  
    \hline 
  $c_{GG}$ & $f_{9}$ & $f_{13}$ & - & - &\\[20pt]   
    \hline 
  $c_{WW}$ & $f_{9}$ & $f_{13}$, $f_{14}$ & - & - &\\[20pt]
    \hline 
  $c_{BB}$ & $f_{9}$ & $f_{13}$, $f_{14}$ & - & - &\\[20pt]
    \hline 
  $c_{WB}$ & $f_{9}$ & $f_{13}$, $f_{14}$ & - & - &\\[20pt]
    \hline 
  $c_{W}$ & - & $f_{15a}$, $f_{15b}$ & - & - &\\[20pt]
    \hline 
  $c_{B}$ & - & $f_{15a}$, $f_{15b}$ & - & - &\\[20pt]
    \hline 
  $c_{D}$ & - & $f_{12c}$ & - & - &\\[20pt]
  \hline 
  \end{tabular}
\end{center}
\caption{For each Wilson coefficient $c_{i}$, this Table shows which universal coefficients $f_{N}$ contribute to the term proportional to $X_{t}^{n}$ for that $c_i$.}
\label{tab:fcontribc}
\end{table}

We remark that our $c_{GG}, c_6, c_R, c_T$ and $c_H$ coefficients are in agreement with Ref.~\cite{RanHuo}, while $c_{WW}, c_{BB}, c_{WB}, c_W, c_B$ and $c_D$ are not. This is the case even after redundancies in the basis of \cite{RanHuo} are taken into account using the appropriate identities, 
\begin{align*}
\mathcal{O}_{HW} = \mathcal{O}_W - \frac{1}{4}(\mathcal{O}_{WW} + \mathcal{O}_{WB}) \, , \\
\mathcal{O}_{HB} = \mathcal{O}_B - \frac{1}{4}(\mathcal{O}_{BB} + \mathcal{O}_{WB}) \, .
\end{align*}
However, we find that our combination of coefficients corresponding to the $S$ parameter~\footnote{We note that the contribution from the universal coefficient $f_{14}$ to $c_W, c_B$ and $c_{WB}$ cancels out and has no effect on the $S$ parameter. We also see that the operator structure of $f_{14}$ can yield directly the redundant $c_{HW}$ and $c_{HB}$ before being eliminated by integration by parts identities. }, 
\begin{equation*}
S = 4\pi v^2 ( 4c_{WB} + c_W + c_B ) \, ,
\end{equation*}
is in agreement with the expression for the $S$ parameter contribution from squarks given in Ref.~\cite{FRW}, which used the calculation of Ref.~\cite{dreesetal}, while the $S$ parameter using the combination of coefficients from \cite{RanHuo} disagrees. As another useful cross-check, all our coefficients coincide with those of \cite{HLMstop,HLM} in the limit of degenerate masses, while those of \cite{RanHuo} recover the same degenerate expressions for all except that of $c_D$, which differs by a factor of 2 from \cite{HLMstop,HLM} and our expression. We have also checked that the coefficient $c_T$ corresponding to the $T$ parameter is the same as that of~\cite{FRW}.

{\small

\begin{eqnarray*}
c_{GG} &=&
 \frac{1}{24} \left( \frac{h_t^2 - \frac{1}{6} g_1^2 c_{2\beta}}{ m_{\tilde{Q}}^2} + \frac{h_t^2 + \frac{1}{3} g_1^2 c_{2\beta}}{ m_{\tilde{t}_R}^2} - \frac{{\bar X_t}^2}{m_{\tilde{Q}}^2   m_{\tilde{t}_R}^2 } \right)	\, ,\nonumber \\
c_{WW}&=& \frac{6 h_t^2-g_{1}^2 c_{2 \beta }}{96 m_{\tilde{Q}}^2}+{\bar X_t}^2 \left(-\frac{\left(m_{\tilde{Q}}^2+m_{\tilde{t}_R}^2\right) \left(-8 m_{\tilde{Q}}^2 m_{\tilde{t}_R}^2+m_{\tilde{Q}}^4+m_{\tilde{t}_R}^4\right)}{16 m_{\tilde{Q}}^2 \left(m_{\tilde{Q}}^2-m_{\tilde{t}_R}^2\right){}^4}-\frac{3 m_{\tilde{Q}}^2 m_{\tilde{t}_R}^4 \ln \left(\frac{m_{\tilde{Q}}^2}{m_{\tilde{t}_R}^2}\right)}{4 \left(m_{\tilde{Q}}^2-m_{\tilde{t}_R}^2\right){}^5}\right) \, , \nonumber \\
c_{BB}&=& \frac{1}{864} \left(\frac{6 h_t^2-g_{1}^2 c_{2 \beta }}{m_{\tilde{Q}}^2}+\frac{32 \left(g_{1}^2 c_{2 \beta }+3 h_t^2\right)}{m_{\tilde{t}_R}^2}\right) \nonumber \\
&& +{\bar X_t}^2 \left(-\frac{-103 m_{\tilde{Q}}^6 m_{\tilde{t}_R}^2-39 m_{\tilde{Q}}^4 m_{\tilde{t}_R}^4+17 m_{\tilde{Q}}^2 m_{\tilde{t}_R}^6+16 m_{\tilde{Q}}^8+m_{\tilde{t}_R}^8}{144 m_{\tilde{Q}}^2 m_{\tilde{t}_R}^2 \left(m_{\tilde{Q}}^2-m_{\tilde{t}_R}^2\right){}^4} + \frac{\left(m_{\tilde{Q}}^2 m_{\tilde{t}_R}^4-4 m_{\tilde{Q}}^4 m_{\tilde{t}_R}^2\right) \ln \left(\frac{m_{\tilde{Q}}^2}{m_{\tilde{t}_R}^2}\right)}{4 \left(m_{\tilde{Q}}^2-m_{\tilde{t}_R}^2\right){}^5} \right) \, , \nonumber \\
c_{WB}&=&  -\frac{g_{2}^2 c_{2\beta}+2 h_{t}^2}{48 m_{\tilde{Q}}^2}  + {\bar X_t}^2 \left(\frac{33 m_{\tilde{Q}}^4 m_{\tilde{t}_R}^2-3 m_{\tilde{Q}}^2 m_{\tilde{t}_R}^4+5 m_{\tilde{Q}}^6+m_{\tilde{t}_R}^6}{24 m_{\tilde{Q}}^2 \left(m_{\tilde{Q}}^2-m_{\tilde{t}_R}^2\right){}^4}-\frac{m_{\tilde{Q}}^2 m_{\tilde{t}_R}^2 \left(2 m_{\tilde{Q}}^2+m_{\tilde{t}_R}^2\right) \ln \left(\frac{m_{\tilde{Q}}^2}{m_{\tilde{t}_R}^2}\right)}{2 \left(m_{\tilde{Q}}^2-m_{\tilde{t}_R}^2\right){}^5}\right ) \, , \nonumber \\
c_{W}&=& {\bar X_t}^2 \left(\frac{-8 m_{\tilde{Q}}^2 m_{\tilde{t}_R}^2+m_{\tilde{Q}}^4-17 m_{\tilde{t}_R}^4}{12 \left(m_{\tilde{Q}}^2-m_{\tilde{t}_R}^2\right){}^4}+\frac{\left(3 m_{\tilde{Q}}^2 m_{\tilde{t}_R}^4+m_{\tilde{t}_R}^6\right) \ln \left(\frac{m_{\tilde{Q}}^2}{m_{\tilde{t}_R}^2}\right)}{2 \left(m_{\tilde{Q}}^2-m_{\tilde{t}_R}^2\right){}^5}\right) \, , \nonumber \\
c_{B}&=& {\bar X_t}^2 \left(\frac{-8 m_{\tilde{Q}}^2 m_{\tilde{t}_R}^2-23 m_{\tilde{Q}}^4+7 m_{\tilde{t}_R}^4}{12 \left(m_{\tilde{Q}}^2-m_{\tilde{t}_R}^2\right){}^4}-\frac{\left(-12 m_{\tilde{Q}}^4 m_{\tilde{t}_R}^2+3 m_{\tilde{Q}}^2 m_{\tilde{t}_R}^4-4 m_{\tilde{Q}}^6+m_{\tilde{t}_R}^6\right) \ln \left(\frac{m_{\tilde{Q}}^2}{m_{\tilde{t}_R}^2}\right)}{6 \left(m_{\tilde{Q}}^2-m_{\tilde{t}_R}^2\right){}^5}\right) \, , \nonumber \\
c_{D}&=& {\bar X_t}^2 \left(\frac{10 m_{\tilde{Q}}^2 m_{\tilde{t}_R}^2+m_{\tilde{Q}}^4+m_{\tilde{t}_R}^4}{2 \left(m_{\tilde{Q}}^2-m_{\tilde{t}_R}^2\right){}^4}-\frac{3 m_{\tilde{Q}}^2 m_{\tilde{t}_R}^2 \left(m_{\tilde{Q}}^2+m_{\tilde{t}_R}^2\right) \ln \left(\frac{m_{\tilde{Q}}^2}{m_{\tilde{t}_R}^2}\right)}{\left(m_{\tilde{Q}}^2-m_{\tilde{t}_R}^2\right){}^5}\right) \, ,\\
c_{H}&=&  \frac{1}{144} \left(\frac{\left(g_1^2 c_{2 \beta }-6 h_t^2\right){}^2}{m_{\tilde{Q}}^2}+\frac{8 \left(g_1^2 c_{2 \beta }+3 h_t^2\right){}^2}{m_{\tilde{t}_R}^2}\right) \nonumber \\
&&  + \frac{{\bar X_t}^{2}}{24 m_{\tilde{Q}}^2 m_{\tilde{t}_R}^2 \left(m_{\tilde{Q}}^2-m_{\tilde{t}_R}^2\right){}^3} \Big( m_{\tilde{Q}}^4 m_{\tilde{t}_R}^2 \left(\left(44 g_1^2-9 g_2^2\right) c_{2 \beta }+78 h_t^2\right)+m_{\tilde{Q}}^2 m_{\tilde{t}_R}^4 \left(\left(26 g_1^2-45 g_2^2\right) c_{2 \beta }-102 h_t^2\right) \nonumber \\
&& -8 m_{\tilde{Q}}^6 \left(g_1^2 c_{2 \beta }+3 h_t^2\right)+2 m_{\tilde{t}_R}^6 \left(6 h_t^2-g_1^2 c_{2 \beta }\right)  \Big) \nonumber \\
&& + \frac{{\bar X_t}^{2} \ln \left(\frac{m_{\tilde{Q}}^2}{m_{\tilde{t}_R}^2}\right) \left(3 m_{\tilde{t}_R}^4 \left(g_2^2 c_{2 \beta }+2 h_t^2\right)+2 \left(3 g_2^2-5 g_1^2\right) c_{2 \beta } m_{\tilde{Q}}^2 m_{\tilde{t}_R}^2\right)}{4 \left(m_{\tilde{Q}}^2-m_{\tilde{t}_R}^2\right){}^4} \, , \nonumber 
\end{eqnarray*}
\begin{eqnarray}
c_{T}&=& \frac{\left(g_2^2 c_{2 \beta }+2 h_t^2\right){}^2}{16 m_{\tilde{Q}}^2} \nonumber \\
&& +{\bar X_t}^2 \left(-\frac{\left(-5 m_{\tilde{Q}}^2 m_{\tilde{t}_R}^2+m_{\tilde{Q}}^4-2 m_{\tilde{t}_R}^4\right) \left(g_2^2 c_{2 \beta }+2 h_t^2\right)}{8 m_{\tilde{Q}}^2 \left(m_{\tilde{Q}}^2-m_{\tilde{t}_R}^2\right){}^3}-\frac{3 m_{\tilde{t}_R}^4 \left(g_2^2 c_{2 \beta }+2 h_t^2\right) \ln \left(\frac{m_{\tilde{Q}}^2}{m_{\tilde{t}_R}^2}\right)}{4 \left(m_{\tilde{Q}}^2-m_{\tilde{t}_R}^2\right){}^4}\right) \nonumber \\
&&+{\bar X_t}^4 \left(\frac{10 m_{\tilde{Q}}^2 m_{\tilde{t}_R}^2+m_{\tilde{Q}}^4+m_{\tilde{t}_R}^4}{4 m_{\tilde{Q}}^2 \left(m_{\tilde{Q}}^2-m_{\tilde{t}_R}^2\right){}^4}+\frac{3 m_{\tilde{t}_R}^2 \left(m_{\tilde{Q}}^2+m_{\tilde{t}_R}^2\right) \ln \left(\frac{m_{\tilde{Q}}^2}{m_{\tilde{t}_R}^2}\right)}{2 \left(m_{\tilde{t}_R}^2-m_{\tilde{Q}}^2\right){}^5}\right)\, , \nonumber \\
c_{R}&=& +\frac{\left(g_2^2 c_{2 \beta }+2 h_t^2\right){}^2}{8 m_{\tilde{Q}}^2} \nonumber \\
&& + {\bar X_t}^2 \Bigg(\frac{m_{\tilde{Q}}^2 m_{\tilde{t}_R}^2 \left(\left(9 g_1^2-5 g_2^2\right) c_{2 \beta }-28 h_t^2\right)+m_{\tilde{Q}}^4 \left(\left(21 g_1^2-5 g_2^2\right) c_{2 \beta }+44 h_t^2\right)+4 m_{\tilde{t}_R}^4 \left(g_2^2 c_{2 \beta }+2 h_t^2\right)}{8 m_{\tilde{Q}}^2 \left(m_{\tilde{Q}}^2-m_{\tilde{t}_R}^2\right){}^3} \nonumber \\
&& -\frac{\ln \left(\frac{m_{\tilde{Q}}^2}{m_{\tilde{t}_R}^2}\right) \left(4 m_{\tilde{Q}}^4 \left(g_1^2 c_{2 \beta }+3 h_t^2\right)+2 \left(5 g_1^2-3 g_2^2\right) c_{2 \beta } m_{\tilde{Q}}^2 m_{\tilde{t}_R}^2+\left(g_1^2+3 g_2^2\right) c_{2 \beta } m_{\tilde{t}_R}^4\right)}{4 \left(m_{\tilde{Q}}^2-m_{\tilde{t}_R}^2\right){}^4}\Bigg) \nonumber \\
&& +{\bar X_t}^4 \left(\frac{-8 m_{\tilde{Q}}^2 m_{\tilde{t}_R}^2-17 m_{\tilde{Q}}^4+m_{\tilde{t}_R}^4}{2 m_{\tilde{Q}}^2 \left(m_{\tilde{Q}}^2-m_{\tilde{t}_R}^2\right){}^4}+\frac{3 m_{\tilde{Q}}^2 \left(m_{\tilde{Q}}^2+3 m_{\tilde{t}_R}^2\right) \ln \left(\frac{m_{\tilde{Q}}^2}{m_{\tilde{t}_R}^2}\right)}{\left(m_{\tilde{Q}}^2-m_{\tilde{t}_R}^2\right){}^5}\right) \, , \nonumber	\\
c_{6}&=&  +\frac{1}{32} \left(-\frac{2 \left(\frac{2}{3} c_{2 \beta } g_1^2+2 h_t^2\right){}^3}{m_{\tilde{t}_R}^2}-\frac{\left(6 h_t^2-c_{2 \beta } g_1^2\right) \left(144 h_t^4-12 c_{2 \beta } \left(g_1^2-9 g_2^2\right) h_t^2+c_{2 \beta }^2 \left(g_1^4+27 g_2^4\right)\right)}{54 m_{\tilde{Q}}^2}\right) \nonumber \\
&&+\frac{ {\bar X_t}^2}{96 m_{\tilde{Q}}^2 m_{\tilde{t}_R}^2 \left(m_{\tilde{Q}}^2-m_{\tilde{t}_R}^2\right){}^2} \Big( 16 \left(c_{2 \beta } g_1^2+3 h_t^2\right){}^2 m_{\tilde{Q}}^4+\frac{3}{2} \Big(11 g_1^4-18 g_2^2 g_1^2+3 g_2^4-192 h_t^4 \nonumber \\
&& -48 c_{2 \beta } \left(g_1^2+g_2^2\right) h_t^2+\cos_{4\beta} \left(11 g_1^4-18 g_2^2 g_1^2+3 g_2^4\right)\Big) m_{\tilde{t}_R}^2 m_{\tilde{Q}}^2+\left(c_{2 \beta } \left(g_1^2-3 g_2^2\right)-12 h_t^2\right){}^2 m_{\tilde{t}_R}^4 \Big) \nonumber \\
&& -\frac{ {\bar X_t}^2 \ln \left(\frac{m_{\tilde{Q}}^2}{m_{\tilde{t}_R}^2}\right) c_{2 \beta } \left(5 g_1^2-3 g_2^2\right) \left(4 \left(c_{2 \beta } g_1^2+3 h_t^2\right) m_{\tilde{Q}}^2+\left(c_{2 \beta } \left(g_1^2-3 g_2^2\right)-12 h_t^2\right) m_{\tilde{t}_R}^2\right)}{48 \left(m_{\tilde{Q}}^2-m_{\tilde{t}_R}^2\right){}^3} \nonumber \\
&&+{\bar X_t}^4 \Bigg( -\frac{4 \left(c_{2 \beta } g_1^2+3 h_t^2\right) m_{\tilde{Q}}^4+5 c_{2 \beta } \left(5 g_1^2-3 g_2^2\right) m_{\tilde{t}_R}^2 m_{\tilde{Q}}^2+\left(c_{2 \beta } \left(g_1^2-3 g_2^2\right)-12 h_t^2\right) m_{\tilde{t}_R}^4}{8 m_{\tilde{Q}}^2 m_{\tilde{t}_R}^2 \left(m_{\tilde{Q}}^2-m_{\tilde{t}_R}^2\right){}^3} \nonumber \\
&& + \frac{3 \ln \left(\frac{m_{\tilde{Q}}^2}{m_{\tilde{t}_R}^2}\right) \left(\left(4 h_t^2+c_{2 \beta } \left(3 g_1^2-g_2^2\right)\right) m_{\tilde{Q}}^2-2 \left(2 h_t^2+c_{2 \beta } \left(g_2^2-g_1^2\right)\right) m_{\tilde{t}_R}^2\right)}{4 \left(m_{\tilde{Q}}^2-m_{\tilde{t}_R}^2\right){}^4}  \Bigg) \nonumber \\
&& + {\bar X_t}^6  \left( \frac{m_{\tilde{Q}}^4+10 m_{\tilde{t}_R}^2 m_{\tilde{Q}}^2+m_{\tilde{t}_R}^4}{2 m_{\tilde{Q}}^2 m_{\tilde{t}_R}^2 \left(m_{\tilde{Q}}^2-m_{\tilde{t}_R}^2\right){}^4} +\frac{3 \ln \left(\frac{m_{\tilde{Q}}^2}{m_{\tilde{t}_R}^2}\right) \left(m_{\tilde{Q}}^2+m_{\tilde{t}_R}^2\right)}{\left(m_{\tilde{Q}}^2-m_{\tilde{t}_R}^2\right){}^5} \right) \, .
\end{eqnarray}

} 

\section{Covariant Derivative Expansion for fermionic fields}
\label{app:fermions}

We give here more details if one wants to integrate out heavy fermionic particles.
We consider the one-loop piece of the Lagrangian involving heavy Dirac fermions, denoted by $\Psi$, having a mass matrix $M$ and interactions with light fields encapsulated by the matrix $W$
\begin{equation}
\mathcal{L}_\text{1-loop}[\phi,\Psi] = \Phi^\dagger(\slashed{P} - M - W[\phi(x)])\Phi \, ,
\end{equation}
where $\slashed{P}=\gamma^{\mu}D_{\mu}$, where the $\gamma^{\mu}$ are the usual gamma matrices.
After integrating out the fermionic fields in the path integral one would get an effective action  which, at one-loop, reads
\begin{equation*}
S_{\text{1-loop}}^{\text{eff}} =  i \text{Tr ln}\left( -\slashed{P} + M + W \right)	\, . 
\end{equation*}
After some simple $\gamma$-matrix algebra manipulations, one can rewrite the previous expression as
\begin{equation*}
S_{\text{1-loop}}^{\text{eff}} =  \frac{i}{2} \text{Tr ln}\left( -P^2 + M^2 -\frac{i}{2}\sigma^{\mu\nu}G^{'}_{\mu\nu} +2M.W + W^2 + [\slashed{P},W] \right)	\, ,
\end{equation*}
where we introduce $\sigma^{\mu\nu} \equiv \frac{i}{2}[\gamma^{\mu},\gamma^{\nu}]$.
This expression for the effective action has exactly the same form as in the scalar case Eq.(\ref{eq:lagrangianUV}) if we define the U matrix as
\begin{equation*}
U = -\frac{i}{2}\sigma^{\mu\nu}G^{'}_{\mu\nu} +2M.W + W^2 + [\slashed{P},W]  \, . 
\end{equation*}
With this specific decomposition of the one-loop effective action for fermions, one can apply the formulae that have been derived in the core of the text. We note that the commutator contains a covariant derivative that vanishes in most renormalisable theories but may be especially important when integrating out effective operators in non-renormalisable theories. We leave the extension of our general calculation to such cases for future studies.

\end{appendices}

 \providecommand{\href}[2]{#2}\begingroup\raggedright

\end{document}